\documentclass{egpubl}
\usepackage{egsr19}
 
 \SpecialIssuePaper         % uncomment for final version of Computer Graphics Forum, special issue
\usepackage[T1]{fontenc}
\usepackage{dfadobe}  

\usepackage{cite}  % comment out for biblatex with backend=biber
\BibtexOrBiblatex
\electronicVersion
\PrintedOrElectronic
\ifpdf \usepackage[pdftex]{graphicx} \pdfcompresslevel=9
\else \usepackage[dvips]{graphicx} \fi

\usepackage{egweblnk} 
\usepackage[utf8]{inputenc}
\usepackage[english]{babel}
\usepackage{amsmath}
\usepackage{amsfonts}
\usepackage{placeins}
\usepackage{amssymb}
\usepackage{hyperref}
\usepackage{array}
\usepackage{graphbox}
\usepackage[normalem]{ulem}
\usepackage{rotating}
\usepackage[dvipsnames]{xcolor}
\usepackage{multirow}

\definecolor{todoCol}{rgb}{1,0.5,0}
\definecolor{GDCol}{rgb}{0,0.9,0.3}
\newcommand{\NEW}  [1] {{#1}}
\newcommand{\REM}  [1] {}

\title[Flexible SVBRDF Capture with a Multi-Image Deep Network]%
      {Flexible SVBRDF Capture with a Multi-Image Deep Network}

\author[V. Deschaintre, M. Aittala, F. Durand, G. Drettakis \& A. Bousseau]
{\parbox{\textwidth}{\centering Valentin Deschaintre$^{1,3}$, Miika Aittala$^{2}$, Fredo Durand$^{2}$, George Drettakis$^{1}$ and Adrien Bousseau$^{1}$ }
        \\
{\parbox{\textwidth}{\centering $^1$
Université Côté d'Azur, Inria \\
         $^2$ MIT CSAIL\\
         $^3$ Optis for Ansys\\
       }
}
}
\begin{document}
\teaser{
\includegraphics[width=\textwidth]{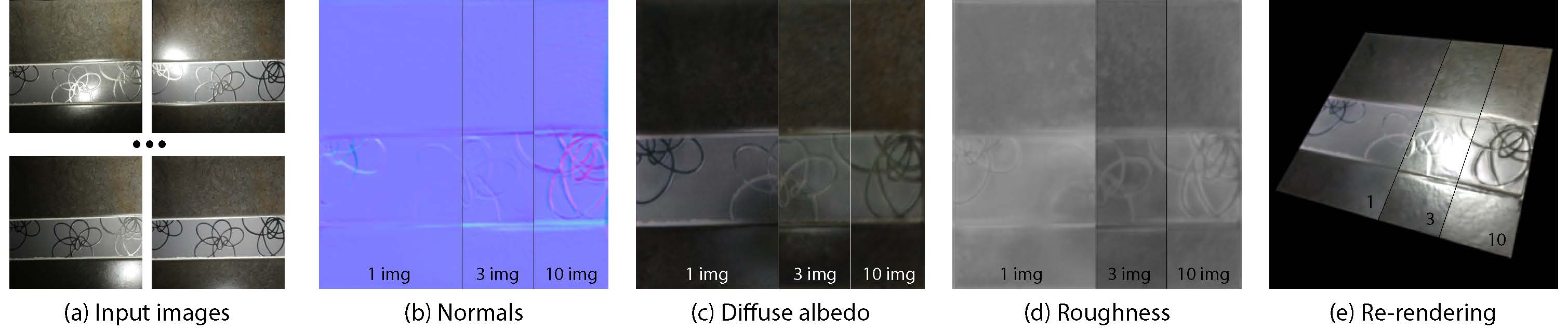}
\caption{Our deep learning method for SVBRDF capture supports a variable number of input photographs taken with \REM{uncontrolled and} uncalibrated light-view directions (a, rectified). While a single image is enough to obtain a first plausible estimate of the SVBRDF maps, more images provide new cues to our method, improving its prediction. In this example, adding images reveals fine normal variations (b), removes highlight residuals in the diffuse albedo (c), and reveals the difference of roughness between the stone, the stripe, and the thin pattern (d). Please see supplemental materials for animated re-renderings.}
\label{fig:teaser}
}

%\begin{teaserfigure}%
%\includegraphics[width=\textwidth]{img/teaser_v0.pdf}
%\vspace{-5mm}
%\caption{From a single flash photograph of a material sample (insets), our deep learning approach predicts a spatially-varying BRDF. See supplemental materials for animations with a moving light.}
%\label{fig:teaser}
%\end{teaserfigure}

\maketitle

\begin{abstract}
	Empowered by deep learning, recent methods for material capture can estimate a spatially-varying reflectance from a single photograph. Such lightweight capture is in stark contrast with the tens or hundreds of pictures required by traditional optimization-based approaches. However, a single image is often simply not enough to observe the rich appearance of real-world materials. We present a deep-learning method capable of estimating material appearance from a variable number of uncalibrated and unordered pictures captured with a handheld camera and flash. Thanks to an order-independent fusing layer, this architecture extracts the most useful information from each picture, while benefiting from strong priors learned from data.  The method can handle both view and light direction variation without calibration.  We show how our method improves its prediction with the number of input pictures, and reaches high quality reconstructions with as little as $1$ to $10$ images -- a sweet spot between existing single-image and complex multi-image approaches.





\end{abstract}

\maketitle

\textcolor{gray}{This paper is a low resolution version of our full paper, available here : \url{https://www-sop.inria.fr/reves/Basilic/2019/DADDB19/}.}  
\section{Introduction}

The appearance of most real-world materials depends on both viewing and lighting directions, which makes their capture a challenging task. 
While early methods achieved faithful capture by densely sampling the view-light conditions \cite{Mcallister2002,dana1999reflectance}, this exhaustive strategy requires expensive and time-consuming hardware setups. 
In contrast, lightweight methods attempt to only perform a few measurements, but require strong prior knowledge on the solution to fill the gaps. In particular, recent methods produce convincing spatially-varying material appearances from a single flash photograph thanks to deep neural networks trained from large quantities of synthetic material renderings \cite{Deschaintre18,Li18}. 
However, in many cases a single photograph simply does not contain enough information to make a good inference for a given material.
Figure~\ref{fig:teaser}(b-d) illustrates typical failure cases of single-image methods, where the flash lighting provides insufficient cues of the relief of the surface, and leaves highlight residuals in the diffuse albedo and specular maps. Only additional pictures with side views or lights reveal fine geometry and reflectance details.

We propose a method that leverages the information provided by additional pictures, while retaining a lightweight capture procedure. 
%In particular, our method accepts a varying number of images, taken with a handheld camera and light. 
When few images are provided, our method harnesses the power of learned priors to make an educated guess, while when additional images are available, our method improves its prediction to best explain all observations. We achieve this flexibility thanks to a deep network architecture capable of processing an arbitrary number of input images with \REM{uncontrolled and} uncalibrated light-view directions. The key observation is that such image sets are fundamentally unstructured. They do not have a meaningful ordering, nor a pre-determined type of content for any given input. Following this reasoning, we adopt a pooling-based network architecture that treats the inputs in a perfectly order-invariant manner, giving it powerful means to extract and combine subtle joint appearance cues scattered across the inputs.

Our flexible approach allows us to capture spatially-varying materials with $1$ to $10$ images, providing a significant improvement
over single-image methods while requiring much fewer images and less constrained capture than traditional multi-image methods.

\section{Related Work}
We first review prior work on appearance capture, focusing on methods working with few images. 
We then discuss deep learning methods capable of processing multiple images.

\paragraph*{Appearance capture.}
The problem of acquiring real-world appearance has been extensively studied in computer graphics and computer vision, as surveyed by Guarnera et al.~\shortcite{Guarnera16}. 
\REM{While} Early efforts focused on capturing appearance under controlled view and lighting conditions\NEW{, first using motorized point lights and cameras \cite{Mcallister2002,dana1999reflectance}
and later using complex light patterns such as linear light sources \cite{Gardner2003}, spherical gradients \cite{ghosh_estimating_2009}, Fourier basis \cite{Aittala13Practical}, or deep-learned patterns \cite{Kang2018}. While these methods provide high-quality capture of complex material effects -- including anisotropy, they require tens to hundreds of measurements acquired using dedicated hardware.
In contrast, }recent work manages to recover plausible spatially-varying appearance (SVBRDF) from very few pictures by leveraging strong priors on natural materials \cite{Wang2011,Aittala15,Aittala16,Ren11,Dong10,Hui2017} and lighting \cite{Lombardi16,Dong14,Riviere17}. In particular, deep learning is nowadays the method of choice to automatically build priors from data, which allows the most recent methods to only use one picture to recover a plausible estimate of the spatially-varying appearance of flat samples \cite{Li17,Ye2018,Deschaintre18,Li18}, and even the geometry of isolated objects \cite{LiShapeRefl:2018}. However, while impressive in many cases, the solutions produced by these single-image methods are largely driven by the learned priors, and often fail to reproduce important material effects simply because they are not observed in the image provided as input, or are too ambiguous to be accurately identified without additional observations. We address this limitation by designing an architecture that supports an arbitrary number of input images. 
Compared to existing single-image methods \cite{Li17,Ye2018,Deschaintre18,Li18}, our multi-image approach produces results of increasing quality as more images are provided. Compared to optimization-based multi-image methods \cite{Riviere16,Hui2017}, our deep-learning approach requires much fewer images to produce high-quality solutions -- $1$ to $10$ instead of around a hundred, while retaining much of the convenience of \REM{uncontrolled} handheld capture. \NEW{Nevertheless, the lightweight nature of our method makes it hard to reach the accuracy of solutions based on calibrated view and light conditions.}
%\TODO{Would be good to provide a comparison to an optimization-based method as supplemental materials, using the gradient descent code in TensorFlow.}

%\paragraph{Training with synthetic data.}
%Because the contributions of shape, material and lighting are conflated in the colors or real-world pictures, many deep-learning methods for inverse rendering rely on synthetic training data to obtain the necessary supervision on these separate components \cite{rematas2017,Li17,Deschaintre18,Li18,LiShapeRefl:2018,liu2017material}. While in theory image synthesis offers the means to generate an arbitrary large amount of training data, the cost of image rendering, storage and transfer limits the size of the datasets used in practice. For example, Li et al.~\shortcite{Li18} and Deschaintre et al.~\shortcite{Deschaintre18} report training datasets of $150{,}000$ and $200{,}000$ images respectively. In contrast, our online data generation allows us to provide the network with a new image at each iteration of the training, yielding up to $2$ million training images in practice. Our online data generation also greatly simplifies testing with different data distributions, a property that we exploit to compare multiple versions of our approach.

\paragraph*{Multi-image deep networks.}
Many computer vision tasks become better posed as the number of observations increases, which calls for methods capable of handling a variable number of input images. For example, classical optimization approaches assign a data fitting error to each observation and minimize their sum. However, implementing an analogous strategy in a deep learning context remains a challenge because most neural network architectures, such as the popular U-Net used in prior work \cite{Li17,Deschaintre18,Li18}, require inputs of a fixed size and treat these inputs in an asymmetric manner. These architectures thus cannot simultaneously benefit from powerful learned priors as well as multiple unstructured observations.

Choy et al. \shortcite{choy20163d} faced this challenge in the context of multi-view 3D reconstruction and proposed a recurrent architecture that processes a sequence of images to progressively refine its prediction. However, the drawback of such an approach is that the solution still depends on the order in which the images are provided to the method -- the first image has a great impact on the overall solution, while subsequent images tend to only modify details. This observation motivated Wiles et al. \shortcite{Wiles2017SilNetS} to process each image of a multi-view set through separate encoders before combining their features through max-pooling, an order-agnostic operation. Aittala et al.~\shortcite{Aittala18} and Chen et al. \shortcite{chen2018ps} apply a similar strategy to the problems of burst image deblurring and photometric stereo, respectively. In the field of geometry processing, Qi et al.~\shortcite{Qi2017} also apply a pooling scheme for deep learning on point sets, and show that such an architecture is an universal approximator for functions whose inputs are set-valued. Zaheer et al.~\shortcite{Zaheer2017} further analyze the theoretical properties of pooling architectures and demonstrate superior performance over recurrent architectures on multiple tasks involving loosely-structured set-valued input data. We build on this family of work to offer a method that processes images captured in an arbitrary order, and that can handle uncalibrated viewing and lighting conditions.

\begin{figure}[!b]
\includegraphics[width=\linewidth]{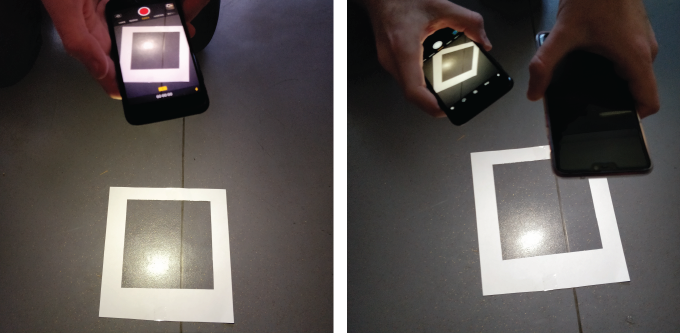}
\vspace{-6mm}
\caption{We use a simple paper frame to help register pictures taken from different viewpoints. We use either a single smartphone and its flash, or two smartphones to cover a larger set of view/light configurations.}
\label{fig:captureSetup}
\end{figure}

\begin{figure*}[t]
\includegraphics[width=\linewidth]{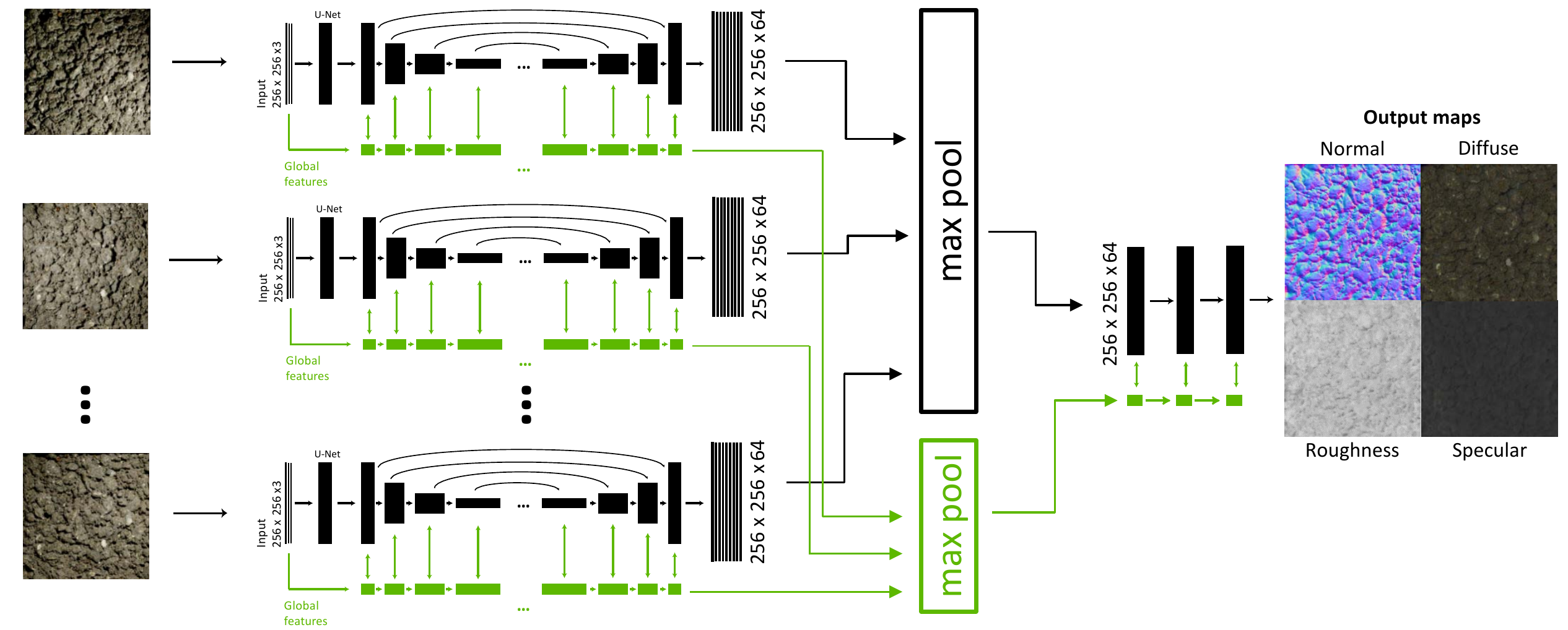}
\caption{Overview of our deep network architecture. Each input image is processed by its copy of the encoder-decoder to produce a feature map. While the number of images and network copies can vary, a pooling layer fuses the output maps to obtain a fixed-size representation of the material, which is then processed by a few convolutional layers to produce the SVBRDF maps.}
\label{fig:netDiagram}
\end{figure*}

\section{Capture Setup}
We designed our method to take as input a variable number of images, captured under \REM{uncontrolled}\NEW{uncalibrated} light and view directions. 
Figure~\ref{fig:captureSetup} shows the capture setup we experimented with, where we place the material sample within a white paper frame and capture it by holding a smartphone in one hand and a flash in the other, or by using the flash of the smartphone as a co-located light source. Similarly to Paterson et al.~\shortcite{Paterson05} and Hui et al.~\shortcite{Hui2017}, we use the four corners of the frame to compute an homography that rectifies the images, and crop the paper pixels away before processing the images with our method. We capture pictures of $3456 \times 3456$ pixels and resize them to $256 \times 256$ pixels after cropping.
%In addition, the mostly diffuse paper pixels provide valuable information about the overall lighting conditions.

\section{Multi-Image Material Inference}
\label{sec:architecture}
Our goal is to estimate the spatially-varying bi-directional reflectance distribution function (SVBRDF) of a flat material sample given a few aligned pictures of that sample. We adopt a parametric representation of the SVBRDF in the form of four maps representing the per-pixel surface normal and diffuse albedo, specular albedo and specular roughness of a Cook-Torrance~\shortcite{Cook82} BRDF model.

The core of our method is a multi-image network composed of several copies of a single-image network , as illustrated in Figure~\ref{fig:netDiagram}.
The number of copies is dynamically chosen to match the number of inputs provided by the user (or the training sample). All copies are identical in their architecture and weights, meaning that each input receives an identical treatment by its respective network copy. 
The findings from each single-image network are then fused by a common order-agnostic pooling layer before being subsequently processed into a joint estimate of the SVBRDF.

We now detail the single-image network and the fusion mechanism, before describing the loss we use to compare the network prediction against a ground-truth SVBRDF. We detail our generation of synthetic training data in Section~\ref{sec:dataset}.

\NEW{The source code of our network architecture along with pre-trained weights is available at \tiny{\url{https://team.inria.fr/graphdeco/projects/multi-materials/}} }

\subsection{Single-image network}
We base our architecture on the single-image network of Deschaintre et al.~\shortcite{Deschaintre18}, which was designed for a similar material acquisition task. The network follows the popular U-Net encoder-decoder architecture \cite{Ronneberger15}, to which it adds a fully-connected track responsible for processing and transmitting global information across distant pixels. While the original architecture outputs four SVBRDF maps, we modify its last layer to instead output a $64$-channel feature map, which retains more information to be processed by the later stages of our architecture. We also provide pixel coordinates as extra channels to the input to help the convolutional network reason about spatial information \cite{Liu2018_Coordconv,Li18}.
%\TODO{Remove if we don't exploit the paper}Yet, we also exploit the paper pixels by \TODO{doing something} and concatenating them to the input of the fully-connected track of the network, since this track was originally designed to process global information about the material at hand.

Since we are targeting a lightweight capture scenario, we do not provide the network with any explicit knowledge of the light and view position. We rather count on the network to deduce related information from visual cues.

%The single-image networks in the first part of our architecture are similar to the one in Deschaintre et al.~\shortcite{Deschaintre18}. It is based on an encoder-decoder architecture with skip connections and a secondary track. he encoder decoder allows to extract the most important high scale \GD{TODO: unclear} information of the input, while the skip connections provide a path for high frequency details to be forwarded through the network. The secondary track is finally added to preserve and reintroduce valuable global information in each layer of the U-net. Rather than directly extracting a material representation, we output \VAR{64} channel maps, leaving enough space to represent the most important information about each input. These representations are then forwarded to the max pooling layer before being processed by the final CNN.

%\section{Network Architecture}

\subsection{Multi-image fusion}
The second part of our architecture fuses the multiple feature maps produced by the single-image networks to form a single feature map of fixed size. 
%Similarly to Aittala et al.~\shortcite{Aittala18} and Chen et al. \shortcite{chen2018ps}, 
%We implement this fusion with a max-pooling layer followed by a sequence of convolutions. 
%\DEL{, motivated by the intuition that this operation preserves the most salient features of each track. }

Specifically, the encoder-decoder track of each single-image network produces a $256 \times 256 \times 64$ intermediate feature map corresponding to the input image it processed. These maps are fused into a single joint feature map of the same size by picking the maximum value reported by any single-image network at each pixel and feature channel. This max-pooling procedure gives every single-image network equal means to contribute to the content of the joint feature map in a perfectly order-independent manner \cite{Aittala18,chen2018ps}.

%\MA{Is it to be believed that the idea of max-pooling is so obvious that it doesn't need to be precisely defined (in particular given the terminology conflict, that it's not the max-pooling you do in typical resolution reduction and so on), and given an intuitive explanation of why a network can be expected to learn to use it meaningfully? Currently the text kind of assumes that the reader will go read the two papers cited and then come back.}\AB{To the contrary, I encourage you to provide a more precise definition and justification of the architecture!}

The pooled intermediate feature map is finally decoded by $3$ layers of convolutions and non-linearities, which provide the network sufficient expressivity to transform the extracted information into four SVBRDF maps. The global features in the fully-connected tracks are max-pooled and decoded in a similar manner.
Through end-to-end training, the single-image networks learn to produce features which are meaningful with respect to the pooling operation and useful for reconstructing the final estimate. 

While we vary the number of copies of the single-view network between $1$ and $5$ during training, an important property of this architecture is that it can process an arbitrarily large number of images during testing because all copies share the same weights, and are ultimately fused by the pooling layer to form a fixed-size feature map. In our experiments, we vary the number of input images from $1$ to $10$ at testing time.

\subsection{Loss}
\label{sec:loss}
%Similarly to recent single-image methods \cite{Deschaintre18,Li18,LiShapeRefl:2018}, 
We evaluate the quality of the network prediction with a differentiable \emph{rendering loss}~\cite{Li18,LiShapeRefl:2018,Deschaintre18}. 
We adopt the loss of Deschaintre et al.~\shortcite{Deschaintre18}, which renders the predicted SVBRDF under multiple light and view directions, and compare these renderings with renderings of the ground-truth SVBRDF under the same conditions. The comparison is performed using an $l_1$ norm on the logarithmic values of the renderings to compress the high dynamic range of specular peaks. 

Following Li et al.~\cite{Li18}, we complement this rendering loss with four $l_1$ losses, each measuring the difference between one of the predicted maps and its ground-truth counterpart. We found this direct supervision to stabilize training. Our final loss is a weighted mixture of all losses, $L = L_{\mathrm{Render}} + 0.1 \big(L_{\mathrm{Normal}} + L_{\mathrm{Diffuse}} + L_{\mathrm{Specular}} + L_{\mathrm{Roughness}}\big)$.

\subsection{Training}
\label{sec:training}

We train our network for 7 days on a Nvidia GTX 1080 TI. We let the training run for 1 million iterations with a batch size of 2 and input sizes of $256 \times 256$ pixels.
%The batch and input size are limited by the amount of memory available of the 1080 TI (11 GB).
We use the Adam optimizer \cite{Kingma14} with a learning rate set to $0.0002$ and $\beta = 0.5$.
%We use dropout in the first layers of the decoder to reduce the risk of over-fitting.

%Maybe add more details about why the different choices of parameters etc...
%Discuss if our training has any specificity, how long, how many images are seen, what GPU, what training order, ....

\section{Online Generation of Training Data}
\label{sec:dataset}

Following prior work on deep-learning for inverse rendering \cite{rematas2017,Li17,Deschaintre18,Li18,LiShapeRefl:2018,liu2017material}, we rely on synthetic data to train our network. 
While in theory image synthesis offers the means to generate an arbitrary large amount of training data, the cost of image rendering, storage and transfer limits the size of the datasets used in practice. For example, Li et al.~\shortcite{Li18} and Deschaintre et al.~\shortcite{Deschaintre18} report training datasets of $150{,}000$ and $200{,}000$ images respectively. This practical challenge motivated us to implement an online renderer that generates a new SVBRDF and its multiple renderings at each iteration of the training, yielding up to $2$ million training images in practice.

We first explain how we generate numerous ground-truth SVBRDFs, before describing the main features of our SVBRDF renderer.

\subsection{SVBRDF synthesis}
\label{svbrdfSynthData}
We rely on procedural, artist-designed SVBRDFs to obtain our training data. Starting from a small set of such SVBRDF maps, Deschaintre et al.~\shortcite{Deschaintre18} perform data augmentation by computing $20{,}000$ convex combinations of random pairs of SVBRDFs. We follow the same strategy, although we implemented this material mixing within TensorFlow~\cite{tensorflow2015}, which allows us to generate a unique SVBRDF for each training iteration while only loading a small set of base SVBRDFs at the beginning of the training process. We use the dataset proposed by Deschaintre et al., which contains $1,850$ SVBRDFs covering common material classes such as plastic, metal, wood, leather, \textit{etc}, all obtained from Allegorithmic Substance Share \cite{Substance}. 
%Finally, we use a simple diffuse white BRDF to simulate a band of paper pixels around the SVBRDF maps.

%Furthermore, many of the input SVBRDFs designed by artists exhibit correlated maps. While such correlation is very frequent, such as in a brick wall where both normals and albedo are discontinuous along the brick edges, some materials like a book cover have a varying albedo but constant normal and shininess. To help the network manage such rare cases, we augment their frequency by randomly flattening the normal and roughness maps for $20\%$ of the SVBRDFs. \TODO{Comment on this based on the outcome of the ablation study}

\subsection{SVBRDF rendering}

We implemented our SVBRDF renderer in TensorFlow, so that it can be called at each iteration of the training process.
Since our network takes rectified images as input, we do not need to simulate perspective projection of the material sample. Instead, our renderer simply takes as input four SVBRDF maps along with a light and view position, and evaluates the resulting rendering equation at each pixel. We augment this basic renderer with several features that simulate common effects encountered in real-world captures:

\paragraph*{Viewing conditions.} We distribute the camera positions over an hemisphere centered on the material sample, and vary its distance by a random amount to allow a casual capture scenario where users may not be able to maintain an exact distance from the target. We also perform random perturbations of the field-of-view (set to $40^\circ$ by default) to simulate different types of cameras. Finally, we apply a random rotation and scaling to the SVBRDF maps before cropping them to $256 \times 256$ pixels, which simulates materials of different orientations and scales.

\paragraph*{Lighting conditions.} We simulate a flash light as a point light with angular fall-off. We again distribute the light positions over an hemisphere at a random distance to simulate a handheld flash. Other random perturbations include the angular fall-off to simulate different types of flash, the light intensity to simulate varying exposure, and the light color to simulate varying white-balance. Finally, we also include the simulation of a surrounding lighting environment in the form of a second light with random position, intensity and color, which is kept fixed for a given input SVBRDF.

\paragraph*{Image post-processing.} We have implemented several common image degradations -- additive Gaussian noise, clipping of radiance values to $1$ to simulate low-dynamic range images, gamma correction and quantization over $8$ bits per channel.
\\

While rendering our training data on the fly incurs additional computation, we found that this overhead is compensated by the time gained in data loading. In our experiments, training our system with online data generation takes approximately as much time as training it with pre-computed data stored on disk, making the actual rendering virtually free.

\section{Results and Evaluation}
We evaluate our method using a test dataset of $32$ ground truth SVBRDFs not present in the set used for training data generation. We also use measured Bidirectional Texture Functions (BTFs) \cite{weinmann2014} to compare the re-renderings of our predictions to real-world appearances. Finally, we used our method to acquire a set of around $80$ real-world materials.
Since our method does not assume a controlled lighting, we used either the camera flash or a separate smartphone as the light source for those acquisitions. 
All results in the figures of the main paper were taken with two phones; please
see supplemental for all results and examples acquired with a single phone. Resulting quality is similar in both cases.

%\GD{specify which for each example (in supplemental too)} \VD{The vast majority of our results are taken using 2 phones, backscattering results only appear in supplemental. We can add a figure}

%While we implemented our method with a multi-view multi-light capture setup, our network architecture can support many other capture scenarios. As an example, we also trained a version of our method with a fixed viewpoint. Figure~\ref{} (green and blue) plots the average RMSE achieved by the two versions, measured on our synthetic test set. Surprisingly, the version with fixed viewpoint performs slightly better. %three instances perform similarly, although the version with fixed viewpoint performs slightly better, while the version without paper performs slightly worse.

%\input{figures/Synth5Improvement/synth5improvement}

\subsection{Number of input images}
\begin{figure}[t]
\includegraphics[width=\linewidth]{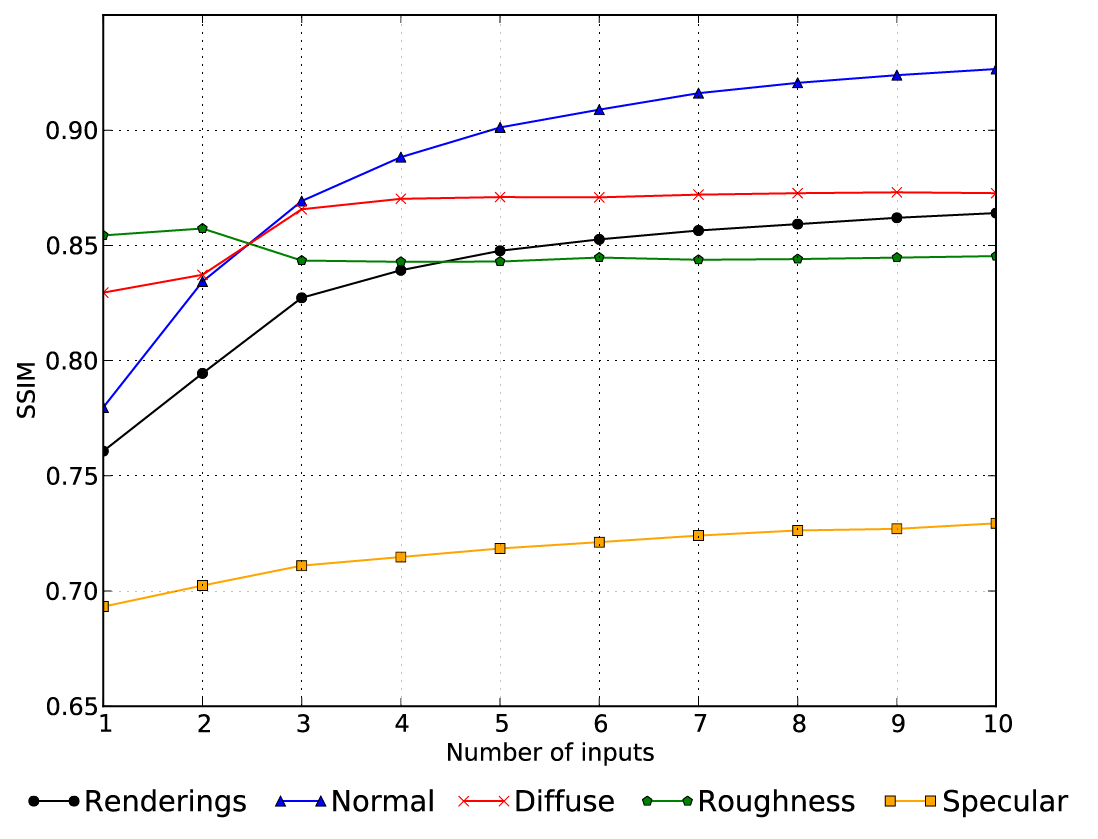}
\caption{SSIM of our predictions with respect to the number of input images, averaged over our synthetic test dataset. 
The SSIM of re-renderings increases quickly for the first images, before stabilizing at around 10 images. The normal maps strongly benefit from new images. Diffuse and specular albedos also improve with additional inputs, which is not the case of the roughness that remains stable overall. We provide similar RMSE plots as supplemental materials.}
\label{fig:globalSSIM}
\end{figure}

\begin{figure}[t]
\includegraphics[width=\linewidth]{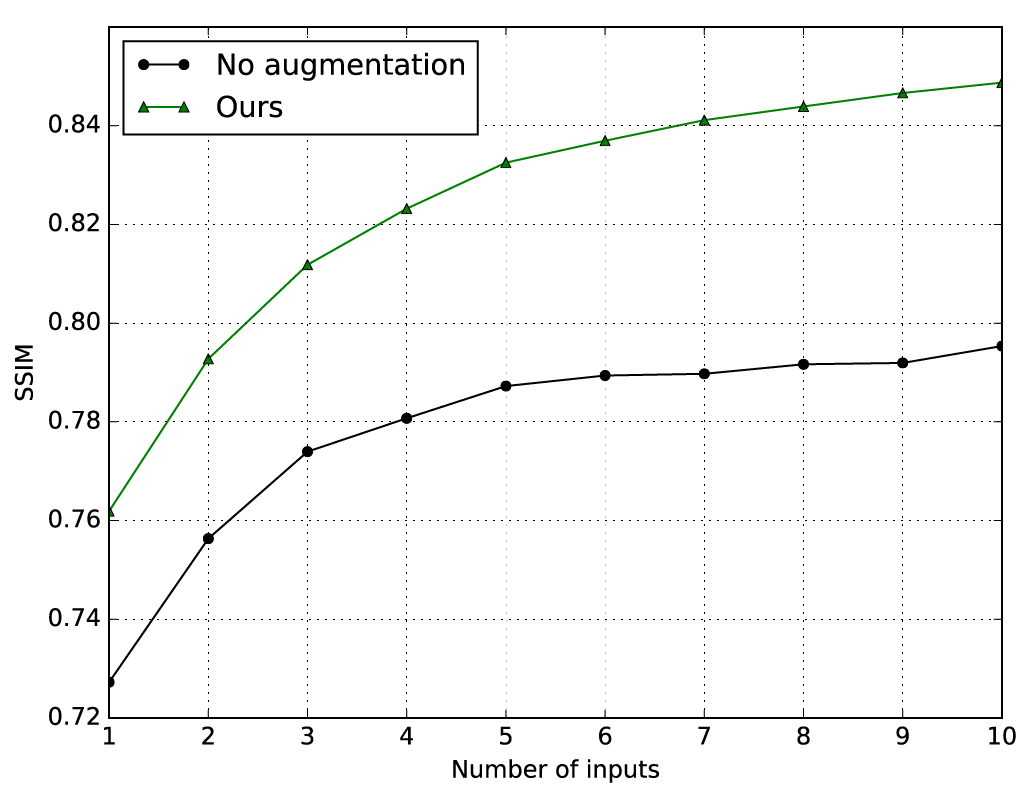}
\vspace{-6mm}
\caption{Ablation study. Comparison of SSIM between our method (green) and a restricted version (black) where the network is trained with lighting and viewing directions chosen 
on a perfect hemisphere, and with all lighting parameters constant (falloff exponent, power, etc.). 
Our complete method achieves higher SSIM when tested on a dataset with small variations of these parameters, showing that it is robust to such perturbations that are frequent in casual real world capture.}
\label{fig:ablation}
\end{figure}

A strength of our method is its ability to cope with a variable number of photographs. 
We first evaluate whether additional images improve the result using synthetic SVBRDFs, for which we have ground truth maps.
%We evaluate this trend quantitatively
We measure the error of our prediction by re-rendering our predicted maps under many views and lights, as done by the rendering loss used for training. Figure~\ref{fig:globalSSIM} plots the SSIM similarity metric of these re-renderings averaged over the test set for an increasing number of images, along with the SSIM of the individual SVBRDF maps. While most improvements happen with the first five images, the similarity continues to increase with subsequent inputs, stabilizing at around 10 images. The diffuse albedo is the fastest to stabilize, consistent with the intuition that few measurements suffice to recover low-frequency signals. Surprisingly, the quality of the roughness prediction seems on average independent of the number of images, suggesting that the method struggles to exploit additional information for this quantity. In contrast, the normal prediction improves with each additional input, as also observed in our experiments with real-world data detailed next. We provide RMSE plots of the same experiment as supplemental materials.

Using the same procedure, in Figure~\ref{fig:ablation} we perform an ablation study to evaluate the impact of including random perturbations of the viewing and lighting conditions in the training data. As expected, the network trained without perturbation does not perform as well as our complete method on our test dataset that includes view and light variations similar to those in casual real world capture. We trained both networks for $750{,}000$ iterations for this experiment.

%We illustrate the effect of using more images wth synthetic ground truth, measured BTFs and qualitative real world captures. We first show a synthetic material, with ground truth available, Figure~\ref{fig:syntheticResults}. We clearly see how normals are significantly improved with more images, and that the characteristic ``splotch'' from specular highlights is mostly removed with the  second input, and disappears completely after \TODO{XXX} images are provided. We also ran numerical tests on the synthetic data. The graph in Figure~\ref{fig:netDiagram} show that on average the relative RMSE decreases with additional inputs. The blue line indicates that using paper in the estimation can deteriorate the results for the synthetic materials. \TODO{Do we want to say this ?}

\begin{figure*}[!t]
\begin{tabular} {ccccccc}
  & \small{Inputs} & \small{Renderings} & \small{Normal} & \small{Diffuse albedo} & \small{Roughness} & \small{Specular albedo} \vspace{1mm} \\
\begin{sideways} \hspace{-5mm} \small{1 input} \end{sideways}  & \hspace{-3mm} \includegraphics[align=c, width=0.145\linewidth]{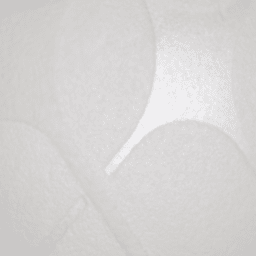} & \hspace{0mm} \includegraphics[align=c, width=0.145\linewidth]{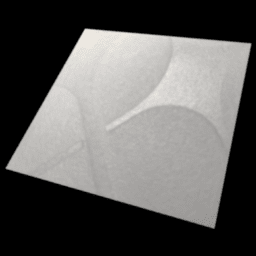} & \hspace{-3mm} \includegraphics[align=c, width=0.145\linewidth]{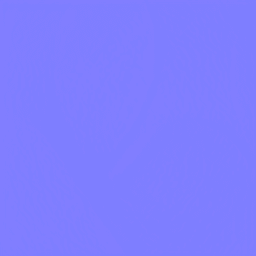} & \hspace{-3mm} \includegraphics[align=c, width=0.145\linewidth]{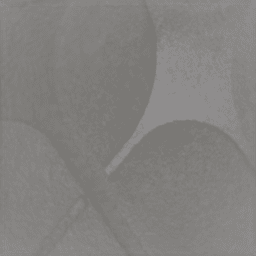} & \hspace{-3mm} \includegraphics[align=c, width=0.145\linewidth]{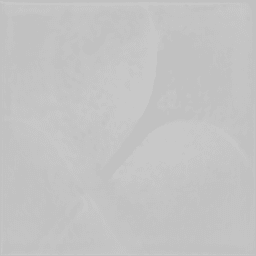} & \hspace{-3mm} \includegraphics[align=c, width=0.145\linewidth]{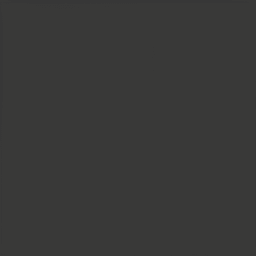} \vspace{1mm} \\
\begin{sideways} \hspace{-5mm} \small{2 inputs} \end{sideways} & \hspace{-3mm} \includegraphics[align=c, width=0.145\linewidth]{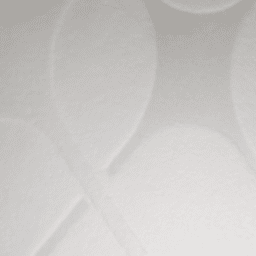} & \hspace{0mm} \includegraphics[align=c, width=0.145\linewidth]{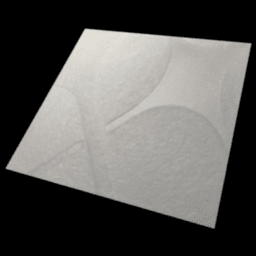} & \hspace{-3mm} \includegraphics[align=c, width=0.145\linewidth]{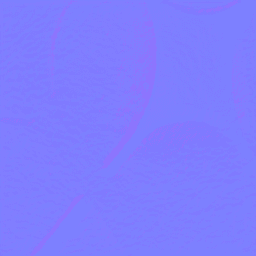} & \hspace{-3mm} \includegraphics[align=c, width=0.145\linewidth]{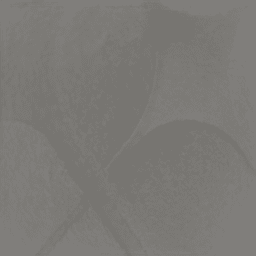} & \hspace{-3mm} \includegraphics[align=c, width=0.145\linewidth]{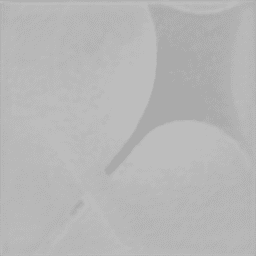} & \hspace{-3mm} \includegraphics[align=c, width=0.145\linewidth]{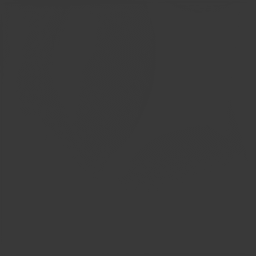} \vspace{1mm} \\
\begin{sideways} \hspace{-5mm} \small{3 inputs} \end{sideways} & \hspace{-3mm} \includegraphics[align=c, width=0.145\linewidth]{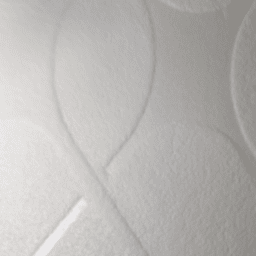} & \hspace{0mm} \includegraphics[align=c, width=0.145\linewidth]{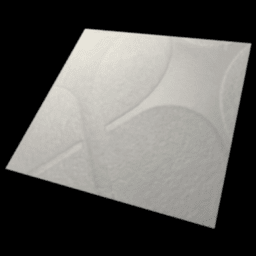} & \hspace{-3mm} \includegraphics[align=c, width=0.145\linewidth]{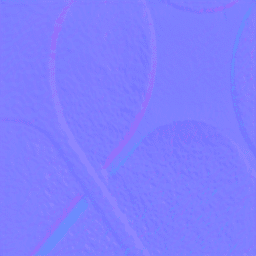} & \hspace{-3mm} \includegraphics[align=c, width=0.145\linewidth]{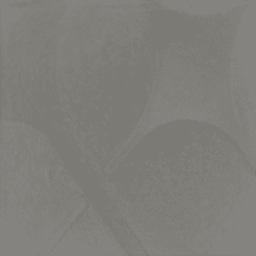} & \hspace{-3mm} \includegraphics[align=c, width=0.145\linewidth]{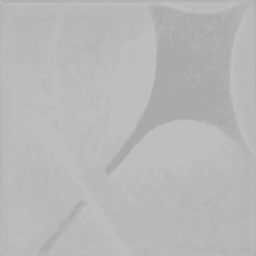} & \hspace{-3mm} \includegraphics[align=c, width=0.145\linewidth]{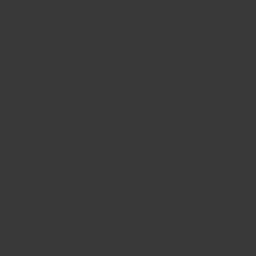} \vspace{1mm} \\
\begin{sideways} \hspace{-5mm} \small{10 inputs} \end{sideways} & \hspace{-3mm} \includegraphics[align=c, width=0.145\linewidth]{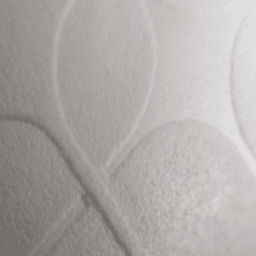} & \hspace{0mm} \includegraphics[align=c, width=0.145\linewidth]{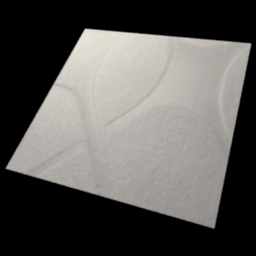} & \hspace{-3mm} \includegraphics[align=c, width=0.145\linewidth]{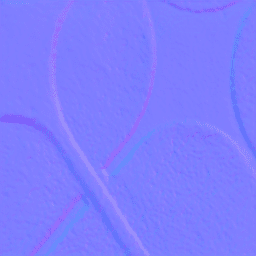} & \hspace{-3mm} \includegraphics[align=c, width=0.145\linewidth]{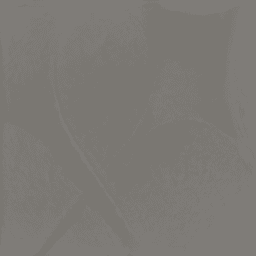} & \hspace{-3mm} \includegraphics[align=c, width=0.145\linewidth]{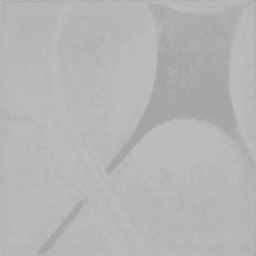} & \hspace{-3mm} \includegraphics[align=c, width=0.145\linewidth]{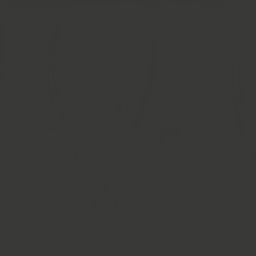} \vspace{1mm} \\
\end{tabular}
\caption{Evaluation on a measured BTF. Three images are enough to capture most of normal and roughness maps. Adding images further improves the result by removing lighting residual from the diffuse albedo, and adding subtle details to the normal and specular maps.}
\label{fig:BonnImageEvolution}
\end{figure*}

Figure~\ref{fig:BonnImageEvolution} shows our predictions on a measured BTF material from the Bonn database \cite{weinmann2014},
using 1, 2, 3 and 10 inputs.
%Most improvements happen with the first three images, but the network does provide progressive improvements with subsequent input, stabilizing at around 10.
For this material, normals, diffuse albedo and roughness estimations improve with more inputs.
In particular, the normal map progressively captures more relief, the diffuse albedo map 
becomes almost uniform, and the embossed part on the upper right is quickly
recognized as shinier than the remaining of the sample.
%The multiple input conditions reveal the complex normal variations, and our network manages to infer that the diffuse albedo is uniform. Similarly, the network needs to see several illumination conditions to determine that the embossed part on the upper right is shinier (less rough).

\begin{figure*}[!t]
\begin{tabular} {ccccccc}
  & \small{Inputs}  & \small{Renderings}  & \small{Normal}  & \small{Diffuse albedo}  & \small{Roughness}  & \small{Specular albedo}  \vspace{1mm} \\
\begin{sideways} \hspace{-5mm} \small{1 input} \end{sideways} & \hspace{-3mm} \includegraphics[align=c, width=0.145\linewidth]{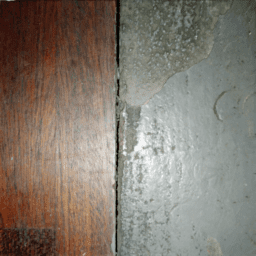} & \hspace{0mm} \includegraphics[align=c, width=0.145\linewidth]{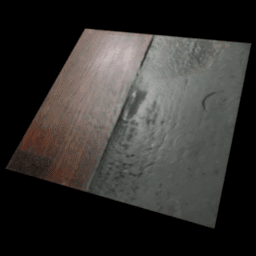} & \hspace{-3mm} \includegraphics[align=c, width=0.145\linewidth]{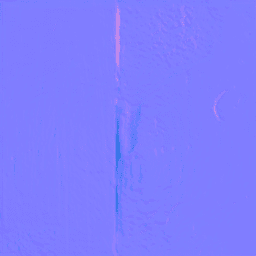} & \hspace{-3mm} \includegraphics[align=c, width=0.145\linewidth]{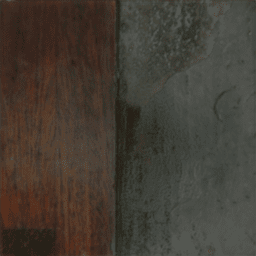} & \hspace{-3mm} \includegraphics[align=c, width=0.145\linewidth]{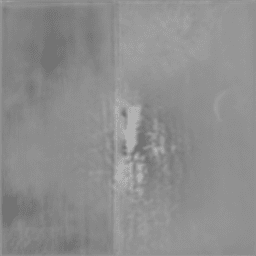} & \hspace{-3mm} \includegraphics[align=c, width=0.145\linewidth]{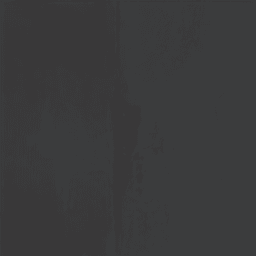} \vspace{1mm} \\
\begin{sideways} \hspace{-5mm} \small{2 inputs} \end{sideways} & \hspace{-3mm} \includegraphics[align=c, width=0.145\linewidth]{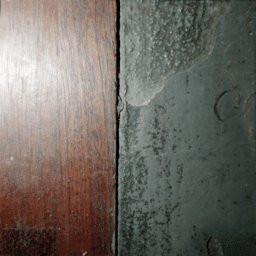} & \hspace{0mm} \includegraphics[align=c, width=0.145\linewidth]{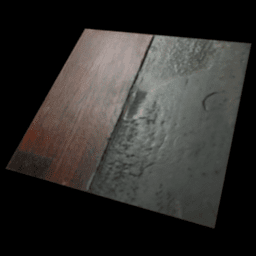} & \hspace{-3mm} \includegraphics[align=c, width=0.145\linewidth]{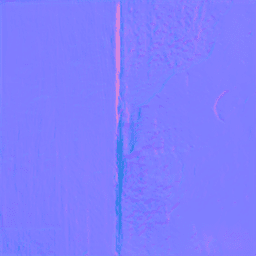} & \hspace{-3mm} \includegraphics[align=c, width=0.145\linewidth]{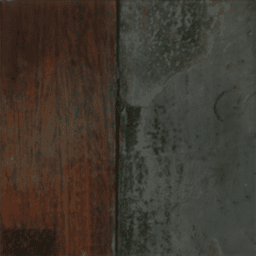} & \hspace{-3mm} \includegraphics[align=c, width=0.145\linewidth]{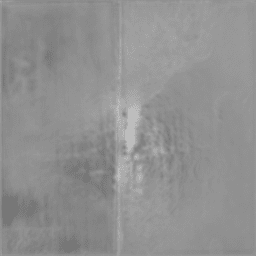} & \hspace{-3mm} \includegraphics[align=c, width=0.145\linewidth]{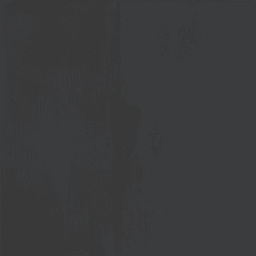} \vspace{1mm} \\
\begin{sideways} \hspace{-5mm} \small{3 inputs} \end{sideways} & \hspace{-3mm} \includegraphics[align=c, width=0.145\linewidth]{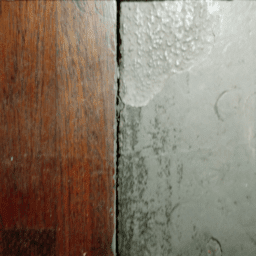} & \hspace{0mm} \includegraphics[align=c, width=0.145\linewidth]{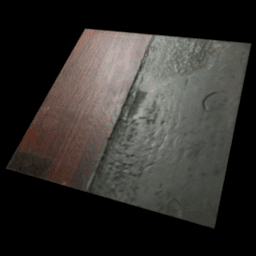} & \hspace{-3mm} \includegraphics[align=c, width=0.145\linewidth]{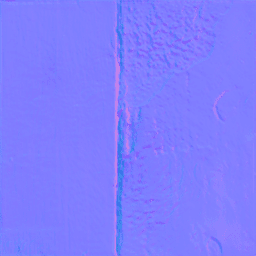} & \hspace{-3mm} \includegraphics[align=c, width=0.145\linewidth]{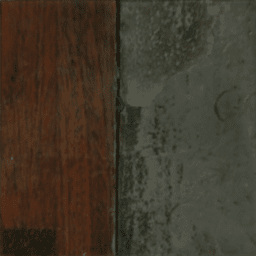} & \hspace{-3mm} \includegraphics[align=c, width=0.145\linewidth]{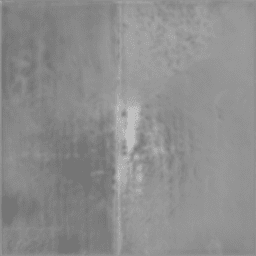} & \hspace{-3mm} \includegraphics[align=c, width=0.145\linewidth]{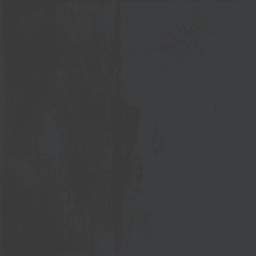} \vspace{1mm} \\
\begin{sideways} \hspace{-5mm} \small{4 inputs} \end{sideways} & \hspace{-3mm} \includegraphics[align=c, width=0.145\linewidth]{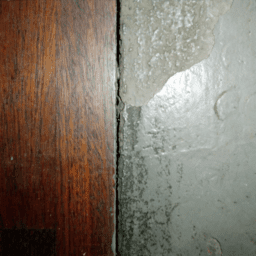} & \hspace{0mm} \includegraphics[align=c, width=0.145\linewidth]{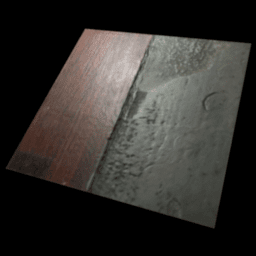} & \hspace{-3mm} \includegraphics[align=c, width=0.145\linewidth]{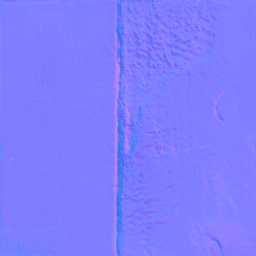} & \hspace{-3mm} \includegraphics[align=c, width=0.145\linewidth]{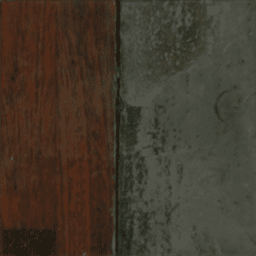} & \hspace{-3mm} \includegraphics[align=c, width=0.145\linewidth]{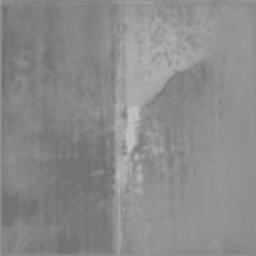} & \hspace{-3mm} \includegraphics[align=c, width=0.145\linewidth]{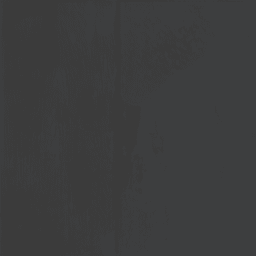} \vspace{1mm} \\
\end{tabular}
\caption{A single flash picture hardly provides enough information for surfaces composed of several materials. In this example, adding images allows the recovery of normal details, and the capture of different roughness values in different parts of the image. Note in particular how the 4th image helps capturing a discontinuity of the roughness on the right part.}
\label{fig:multipleMaterials}
\end{figure*}

For a real material capture we performed (Figure~\ref{fig:multipleMaterials}), we
see similar effects: normals are improved with more inputs, 
%and the specular residuals in the diffuse albedo are progressively removed.
and the difference of roughness between different parts is progressively recovered.
However, we do not have access to ground truth maps for these real-world captures.

Overall, our results in Fig.~\ref{fig:globalSSIM}-\ref{fig:comparisonOneImageReal} and in supplemental material illustrate that our method achieves our goals: adding more pictures greatly improves the results, notably removing artifacts in the diffuse albedo while improving normal estimation. Our method enhances the quality of recovered materials while maintaining a casual capture.

\subsection{Comparison to multi-image optimization}
We compare our data-driven approach to a traditional optimization that takes as input multiple images captured
under the assumption of known and precisely calibrated 
light and viewing conditions. 
Given these conditions 
%view directions, light directions and light power, 
we solve for the SVBRDF maps that minimize re-rendering error
of the input images, as measured by our rendering loss. We further regularize this optimization by augmenting the loss
with a total-variation term that favors piecewise-smooth maps. We solve the optimization with the Adam algorithm \cite{Kingma14}. While the optimization stabilizes after $900$K iterations, we let it run for a total of 2M iterations to ensure full convergence, which takes approximately $3.5$ hours on an NVIDIA GTX 1080 TI.
Given the non-convex nature of the optimization, we initialize the solution to a plausible estimate obtained by setting the diffuse albedo map to the most fronto-parallel input,
the normal map to a constant vector pointing upward, the roughness to zero and the specular albedo to gray.
We use synthetic data for this experiment, which provides us with full control and knowledge of the viewing and lighting conditions needed by the optimization,
as well as with ground truth maps to evaluate the quality of the outcome.

\begin{figure}[t]
\includegraphics[width=\linewidth]{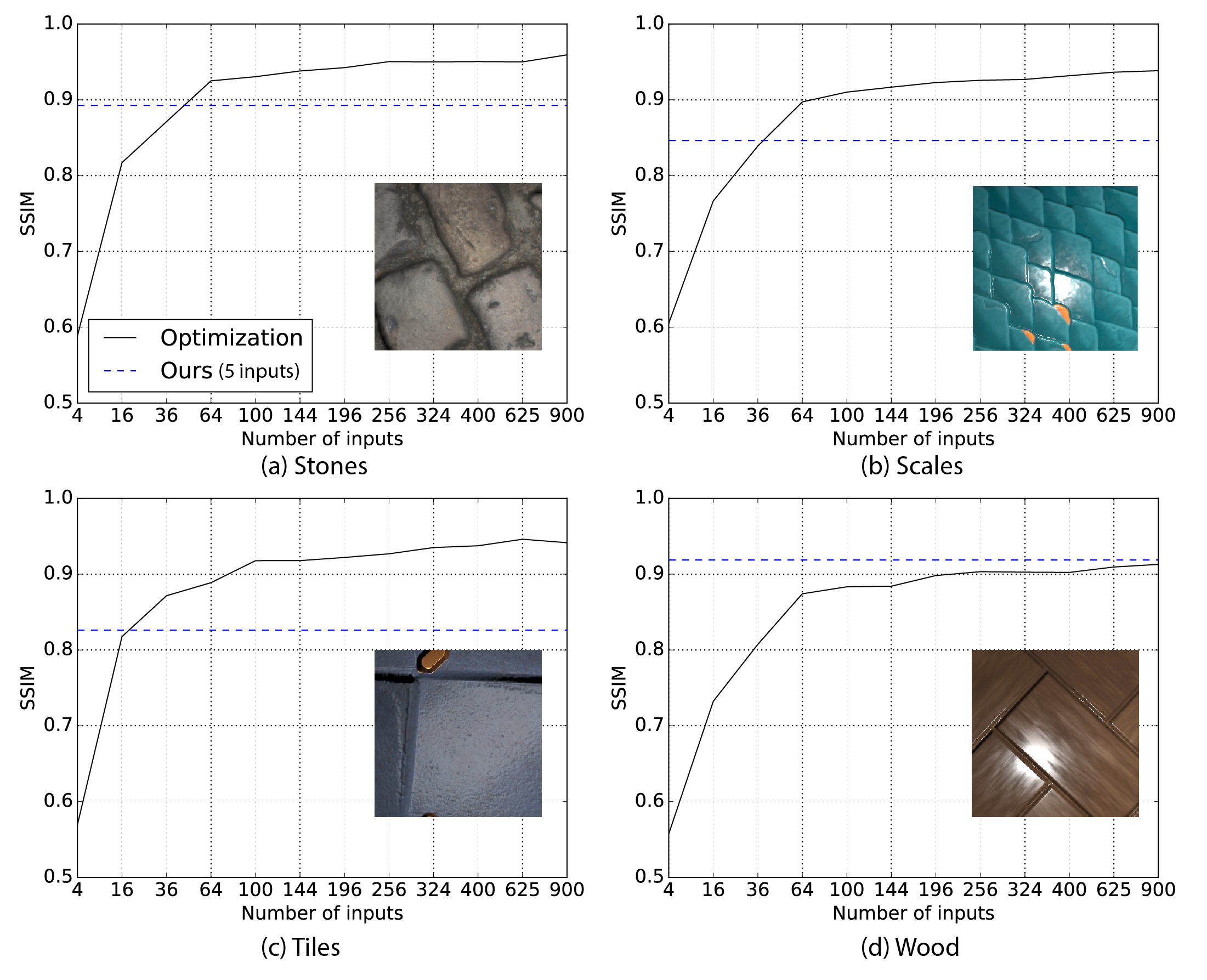}
\vspace{-6mm}
\caption{SSIM on re-renderings for the maps obtained by our method 
with 5 images (dotted blue) and by a classical optimization method 
with an increasing number of input images (black). The classical optimization
requires several dozens of calibrated pictures to outperform our method on rather diffuse
or uniform materials (stones, tiles, scales), while requiring many more for
a more complex material (wood).}
\label{fig:classicalOptim}
\end{figure}

Figure~\ref{fig:classicalOptim} compares the number of input images required to achieve similar quality between the classical optimization and our method,
using view and light directions uniformly distributed over the hemisphere. 
On rather diffuse materials (stones, tiles), the optimization needs a few dozen calibrated images to achieve a result of similar quality to the one produced by our method using only $5$, uncalibrated images. 
A similar number of images is necessary for a material with uniform shininess (scales). However more than $900$ images were necessary for our optimization to reach the quality obtained by our method
on a material with significant normal and roughness variations (wood).
Overall, our method achieves plausible results with much fewer inputs captured under unknown lighting, although classical optimization can recover more precise SVBRDFs if provided with enough carefully-calibrated images.

%The optimization takes an arbitrary number of images and the associated view/light parameters. At each step, it optimizes for parameters (Diffuse albedo, Normal, Roughness and Specular albedo), so that when rendered they provide a result similar to the matching input.
%Add something about the distribution of input images we use maybe ? So that they don't think we use crazy light/view directions. Also discuss the fact that we remove the 5 pixels on the border for the optim result.
%As the optimization is non-convex, it requires careful initialization. We initialize the normal map to point towards z direction, the diffuse map to a fronto parallel picture, Roughness to black and finally specular to gray. We optimize the normal map for two channels, X and Y and derive Z through normalization.

%The optimization function is defined as follow : 
%\\
%$\frac{1}{K} \sum_{i=1}^{K}(|log(I_i + 0.01) - log(O_i + 0.01)| + 0.5 * TV(O_i))$
%\\
%\ref{fig:classicalOptim} compares the number of input images required to achieve similar quality between the classical optimization and our method.  We show that with 5 random images our method achieves similar quality to a classical optimization using \VAR{64} calibrated images.
%While the classical optimization still has the advantage when a large number of calibrated images are provided, our method significantly improve the capture quality using lighweight acquisition scenario.

\subsection{Comparison to alternative deep learning methods}

We first compare our architecture to a simple baseline composed of the network by Deschaintre et al.~\shortcite{Deschaintre18} augmented to take 5 images instead of one. This baseline achieves an average SSIM of $0.826$, similar to the SSIM of $0.847$ produced by our method for the same number of inputs. This evaluation demonstrates that our multi-image network performs as well as a fixed network while providing the freedom to vary the number of input images. 
%RMSE are for ours : 0.0748 and for baseline : 0.0762

%In addition to imposing a fixed number of images, this architecture lacks the ability to treat the 5 inputs in an order-agnostic manner. As a result, when trained with random view and light directions as ours, this baseline achieve an average RMSE of \TODO{how many?}, much higher than the error of \TODO{how many?} produced by our method for the same number of input.

%We compare to a simple solution with 5 random view and light positions, where these are provided together as input to the network. This is shown as the black point in the graph in Fig.~\ref{fig:netDiagram}, and clearly performs worse than our solution.
 
%\input{figures/stats/overall_stats_global_absolute}

%\TODO{Comparison to baseline that would take a fixed number of images, with random lighting}

%\TODO{Comparisons to single-image methods \cite{Deschaintre18,Li18}}

\begin{figure*}[!t]
 \hspace{-7mm}\begin{tabular} {cccccccc}
  & \small{Inputs} & \multicolumn{2}{c}{\small{Renderings}} & \small{Normal} & \small{Diffuse} & \small{Roughness} & \small{Specular}  \vspace{1mm} \\
\begin{sideways} \hspace{-7mm} \tiny{Deschaintre et al. 2018} \end{sideways} & \hspace{-3mm} \includegraphics[align=c, width=0.135\linewidth]{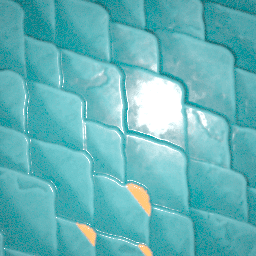} & \hspace{-2.5mm} \includegraphics[align=c, width=0.135\linewidth]{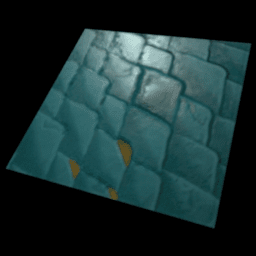} & \hspace{-3mm} \includegraphics[align=c, width=0.135\linewidth]{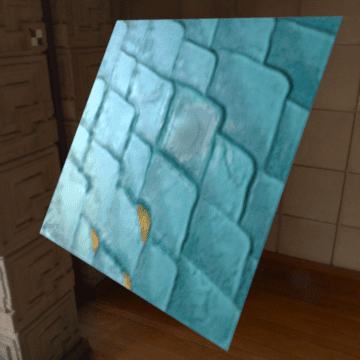} & \hspace{-3mm} \includegraphics[align=c, width=0.135\linewidth]{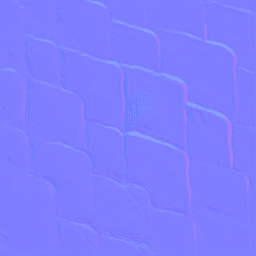} & \hspace{-3mm} \includegraphics[align=c, width=0.135\linewidth]{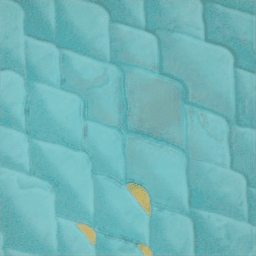} & \hspace{-3mm} \includegraphics[align=c, width=0.135\linewidth]{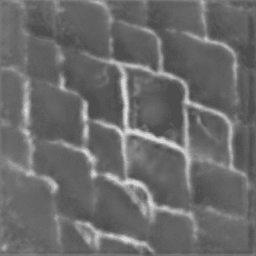} & \hspace{-3mm} \includegraphics[align=c, width=0.135\linewidth]{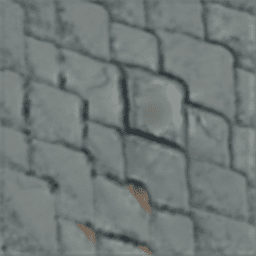} \vspace{1mm} \\
\begin{sideways} \hspace{-7mm} \tiny{Li et al. 2018} \end{sideways} & \hspace{-3mm} \includegraphics[align=c, width=0.135\linewidth]{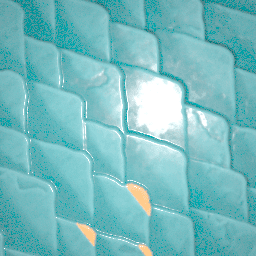} & \hspace{-2.5mm} \includegraphics[align=c, width=0.135\linewidth]{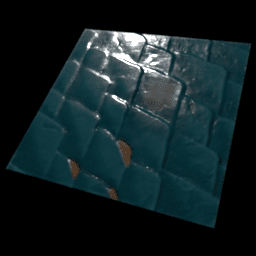} & \hspace{-3mm} \includegraphics[align=c, width=0.135\linewidth]{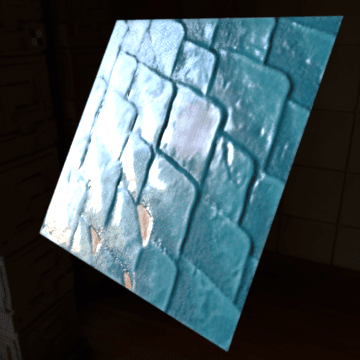} & \hspace{-3mm} \includegraphics[align=c, width=0.135\linewidth]{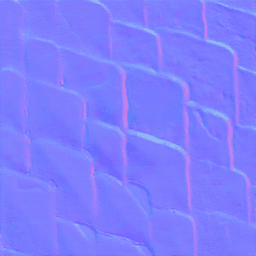} & \hspace{-3mm} \includegraphics[align=c, width=0.135\linewidth]{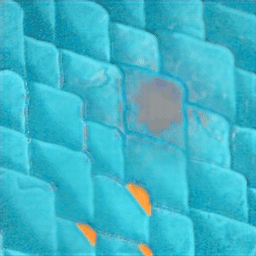} & \hspace{-3mm} \includegraphics[align=c, width=0.135\linewidth]{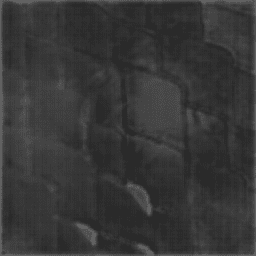} &   \vspace{1mm} \\
\begin{sideways} \hspace{-7mm} \tiny{Ground Truth} \end{sideways} &   & \hspace{-2.5mm} \includegraphics[align=c, width=0.135\linewidth]{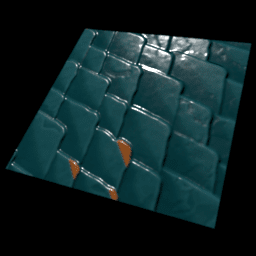} & \hspace{-3mm} \includegraphics[align=c, width=0.135\linewidth]{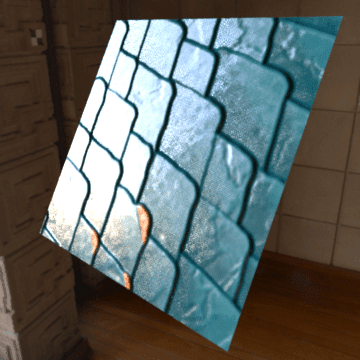} & \hspace{-3mm} \includegraphics[align=c, width=0.135\linewidth]{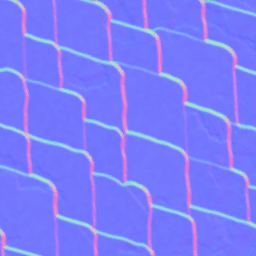} & \hspace{-3mm} \includegraphics[align=c, width=0.135\linewidth]{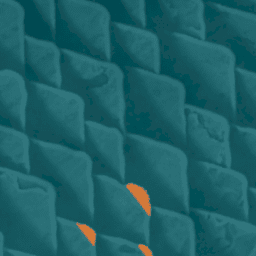} & \hspace{-3mm} \includegraphics[align=c, width=0.135\linewidth]{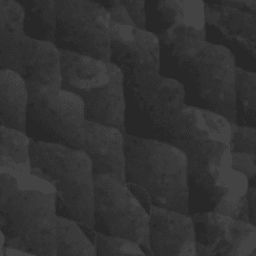} & \hspace{-3mm} \includegraphics[align=c, width=0.135\linewidth]{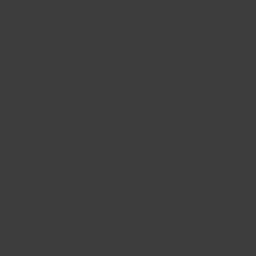} \vspace{1mm} \\
\begin{sideways} \hspace{-7mm} \small{Ours (4 inputs)} \end{sideways} & \begin{tabular} {cc}
\hspace{-2.1 mm} \includegraphics[align=c, width=0.063\linewidth]{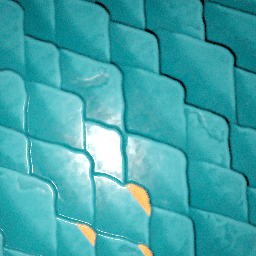} & \hspace{-3.5 mm} \includegraphics[align=c, width=0.063\linewidth]{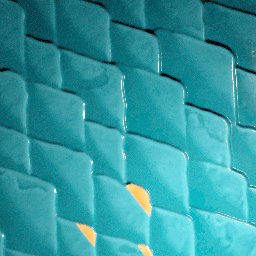} \vspace{1mm} \\
\hspace{-2.1 mm} \includegraphics[align=c, width=0.063\linewidth]{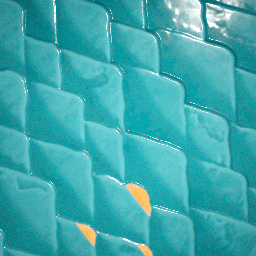} & \hspace{-3.5 mm} \includegraphics[align=c, width=0.063\linewidth]{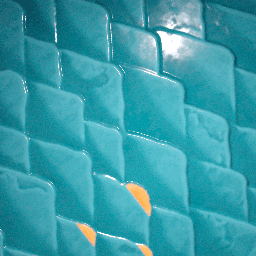} \vspace{1mm} \\
\end{tabular} & \hspace{-2.5mm} \includegraphics[align=c, width=0.135\linewidth]{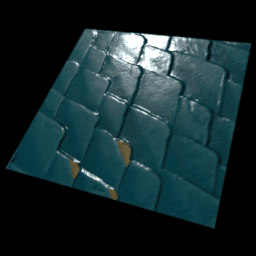} & \hspace{-3mm} \includegraphics[align=c, width=0.135\linewidth]{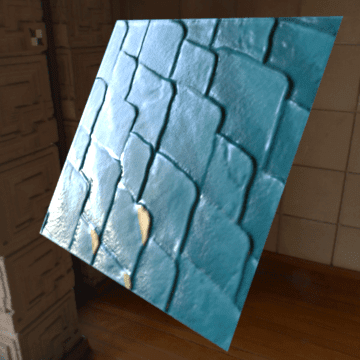} & \hspace{-3mm} \includegraphics[align=c, width=0.135\linewidth]{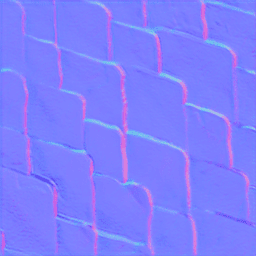} & \hspace{-3mm} \includegraphics[align=c, width=0.135\linewidth]{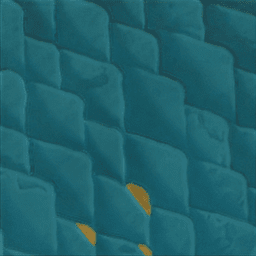} & \hspace{-3mm} \includegraphics[align=c, width=0.135\linewidth]{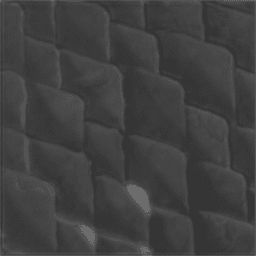} & \hspace{-3mm} \includegraphics[align=c, width=0.135\linewidth]{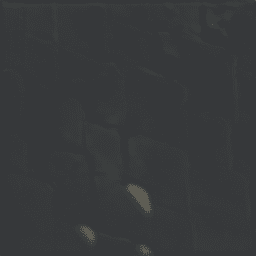} \vspace{3mm} \\
\begin{sideways} \hspace{-7mm} \tiny{Deschaintre et al. 2018} \end{sideways} & \hspace{-3mm} \includegraphics[align=c, width=0.135\linewidth]{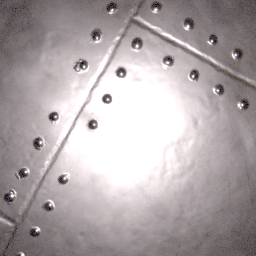} & \hspace{-2.5mm} \includegraphics[align=c, width=0.135\linewidth]{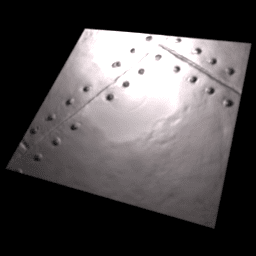} & \hspace{-3mm} \includegraphics[align=c, width=0.135\linewidth]{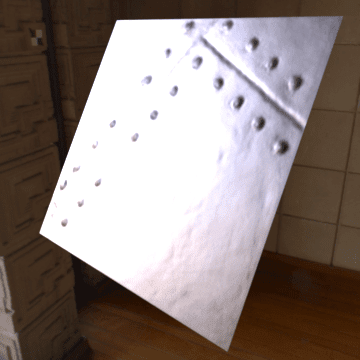} & \hspace{-3mm} \includegraphics[align=c, width=0.135\linewidth]{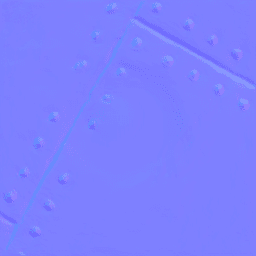} & \hspace{-3mm} \includegraphics[align=c, width=0.135\linewidth]{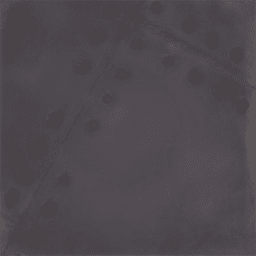} & \hspace{-3mm} \includegraphics[align=c, width=0.135\linewidth]{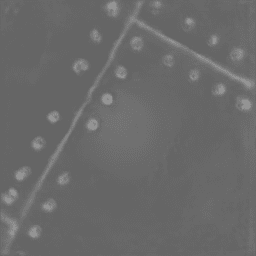} & \hspace{-3mm} \includegraphics[align=c, width=0.135\linewidth]{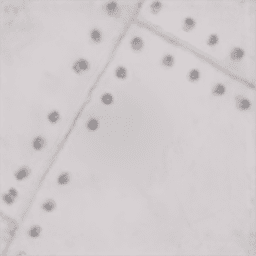} \vspace{1mm} \\
\begin{sideways} \hspace{-7mm} \tiny{Li et al. 2018} \end{sideways} & \hspace{-3mm} \includegraphics[align=c, width=0.135\linewidth]{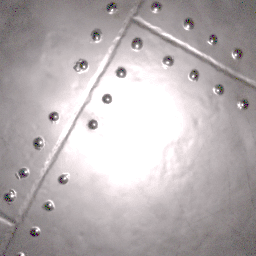} & \hspace{-2.5mm} \includegraphics[align=c, width=0.135\linewidth]{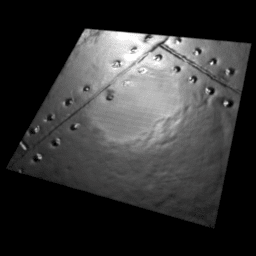} & \hspace{-3mm} \includegraphics[align=c, width=0.135\linewidth]{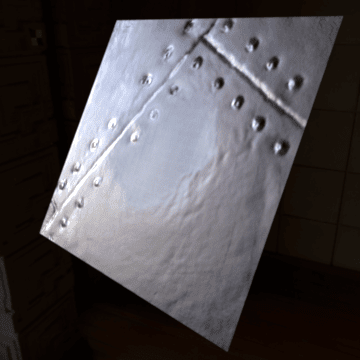} & \hspace{-3mm} \includegraphics[align=c, width=0.135\linewidth]{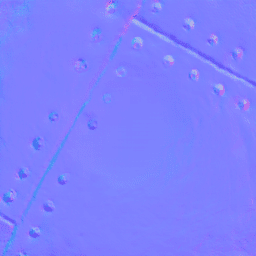} & \hspace{-3mm} \includegraphics[align=c, width=0.135\linewidth]{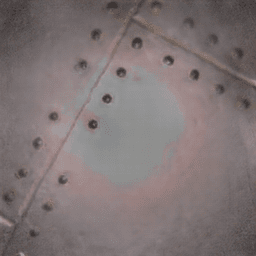} & \hspace{-3mm} \includegraphics[align=c, width=0.135\linewidth]{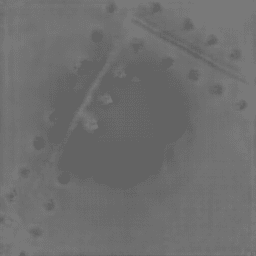} &   \vspace{1mm} \\
\begin{sideways} \hspace{-7mm} \tiny{Ground Truth} \end{sideways} &   & \hspace{-2.5mm} \includegraphics[align=c, width=0.135\linewidth]{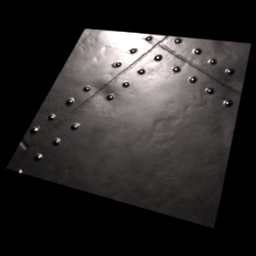} & \hspace{-3mm} \includegraphics[align=c, width=0.135\linewidth]{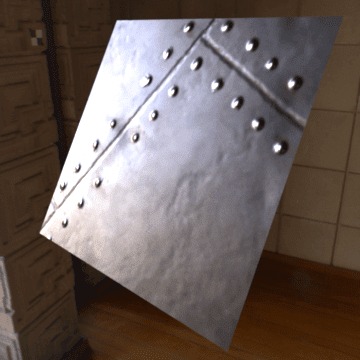} & \hspace{-3mm} \includegraphics[align=c, width=0.135\linewidth]{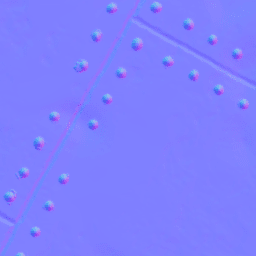} & \hspace{-3mm} \includegraphics[align=c, width=0.135\linewidth]{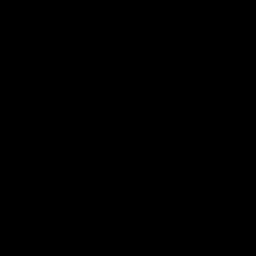} & \hspace{-3mm} \includegraphics[align=c, width=0.135\linewidth]{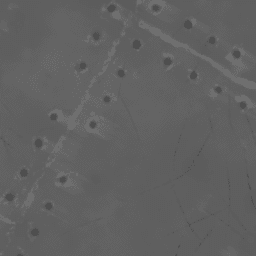} & \hspace{-3mm} \includegraphics[align=c, width=0.135\linewidth]{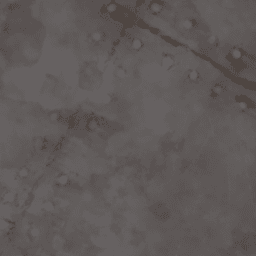} \vspace{1mm} \\
\begin{sideways} \hspace{-7mm} \small{Ours (4 inputs)} \end{sideways} & \begin{tabular} {cc}
\hspace{-2.1 mm} \includegraphics[align=c, width=0.063\linewidth]{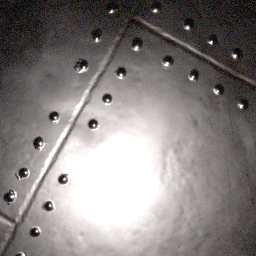} & \hspace{-3.5 mm} \includegraphics[align=c, width=0.063\linewidth]{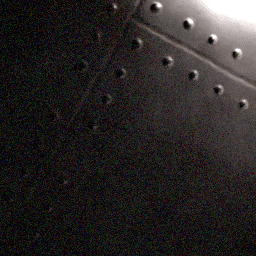} \vspace{1mm} \\
\hspace{-2.1 mm} \includegraphics[align=c, width=0.063\linewidth]{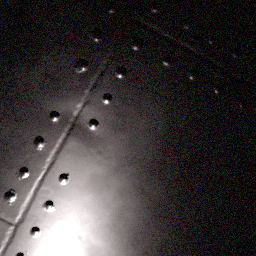} & \hspace{-3.5 mm} \includegraphics[align=c, width=0.063\linewidth]{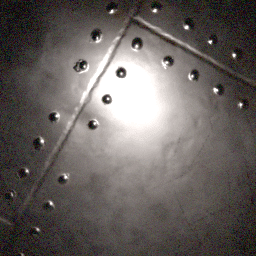} \vspace{1mm} \\
\end{tabular} & \hspace{-2.5mm} \includegraphics[align=c, width=0.135\linewidth]{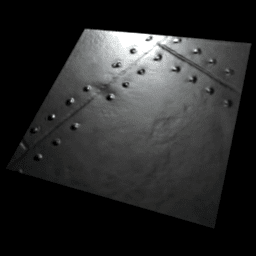} & \hspace{-3mm} \includegraphics[align=c, width=0.135\linewidth]{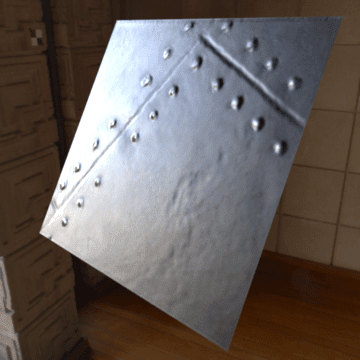} & \hspace{-3mm} \includegraphics[align=c, width=0.135\linewidth]{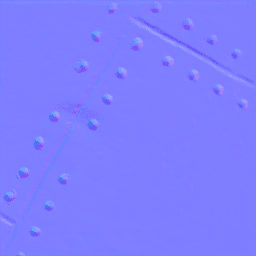} & \hspace{-3mm} \includegraphics[align=c, width=0.135\linewidth]{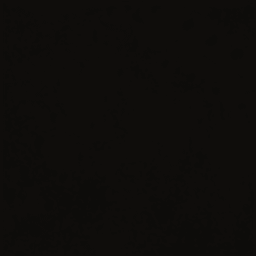} & \hspace{-3mm} \includegraphics[align=c, width=0.135\linewidth]{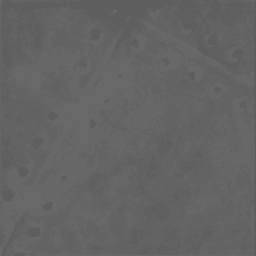} & \hspace{-3mm} \includegraphics[align=c, width=0.135\linewidth]{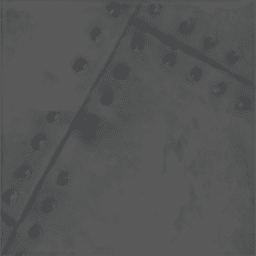} \vspace{1mm} \\
\end{tabular}

\caption{Comparison against single-image methods on synthetic SVBRDFs. Our method leverages additional input images to obtain SVBRDF maps closer to ground truth. In particular, single-image methods under-estimate normal variations and fail to remove the saturated highlight on shiny materials. See supplemental materials for more comparisons and results.}
\label{fig:comparisonOneImageSynth}
\end{figure*}

\begin{figure*}[!t]
\hspace{-7mm}\begin{tabular} {ccccccccc}
  & \small{Inputs} & \multicolumn{2}{c}{\small{Renderings}} & \small{Normal} & \small{Diffuse} & \small{Roughness} & \small{Specular} & 
 \vspace{1mm} \\
\begin{sideways} \hspace{-7mm} \tiny{Deschaintre et al. 2018} \end{sideways} & \hspace{-3mm} \includegraphics[align=c, width=0.135\linewidth]{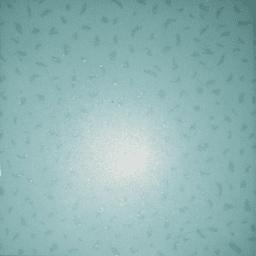} & \hspace{-2.5mm} \includegraphics[align=c, width=0.135\linewidth]{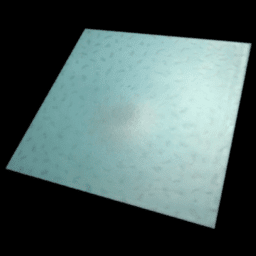} & \hspace{-3mm} \includegraphics[align=c, width=0.135\linewidth]{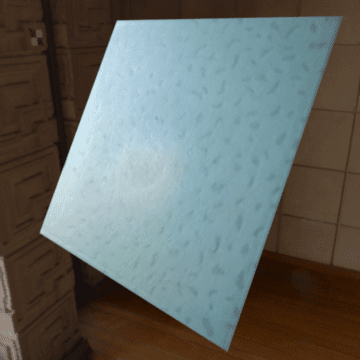} & \hspace{-3mm} \includegraphics[align=c, width=0.135\linewidth]{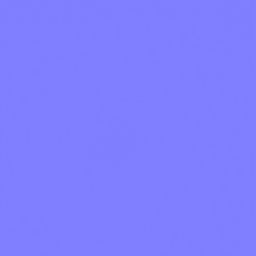} & \hspace{-3mm} \includegraphics[align=c, width=0.135\linewidth]{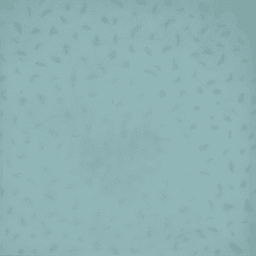} & \hspace{-3mm} \includegraphics[align=c, width=0.135\linewidth]{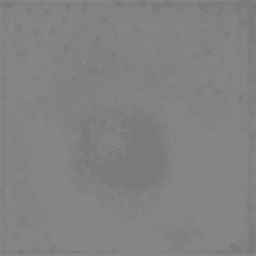} & \hspace{-3mm} \includegraphics[align=c, width=0.135\linewidth]{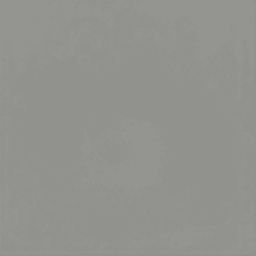} \vspace{1mm} \\
\begin{sideways} \hspace{-7mm} \tiny{Li et al. 2018} \end{sideways} & \hspace{-3mm} \includegraphics[align=c, width=0.135\linewidth]{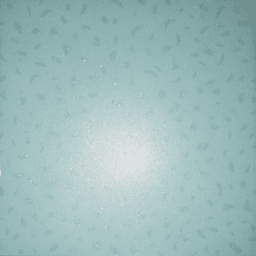} & \hspace{-2.5mm} \includegraphics[align=c, width=0.135\linewidth]{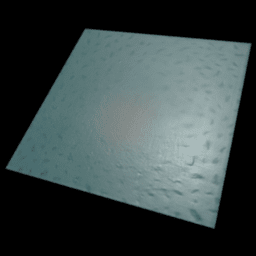} & \hspace{-3mm} \includegraphics[align=c, width=0.135\linewidth]{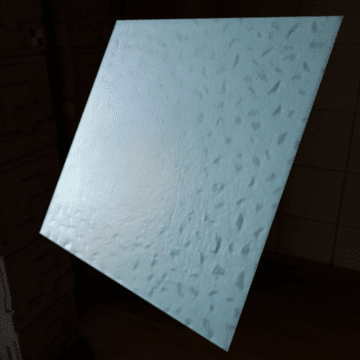} & \hspace{-3mm} \includegraphics[align=c, width=0.135\linewidth]{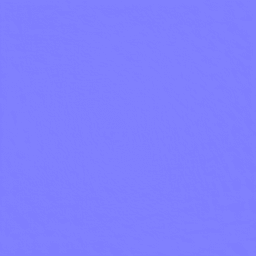} & \hspace{-3mm} \includegraphics[align=c, width=0.135\linewidth]{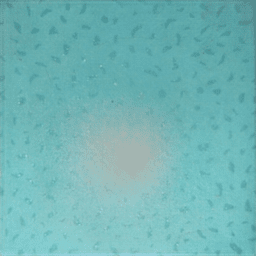} & \hspace{-3mm} \includegraphics[align=c, width=0.135\linewidth]{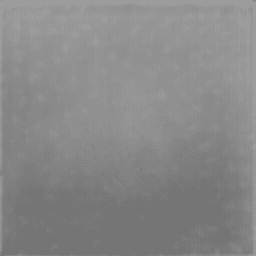} &   \vspace{1mm} \\
\begin{sideways} \hspace{-7mm} \small{Ours (4 inputs)} \end{sideways} & \begin{tabular} {cc}
\hspace{-2.5mm} \includegraphics[align=c, width=0.063\linewidth]{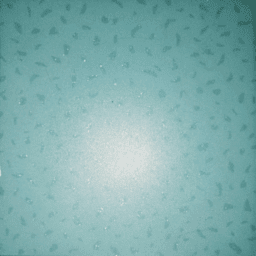} & \hspace{-3mm} \includegraphics[align=c, width=0.063\linewidth]{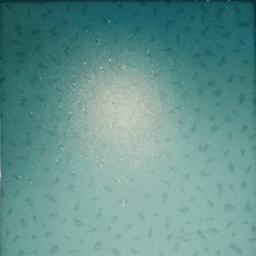} \vspace{1mm} \\
\hspace{-2.5mm} \includegraphics[align=c, width=0.063\linewidth]{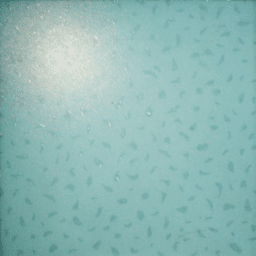} & \hspace{-3mm} \includegraphics[align=c, width=0.063\linewidth]{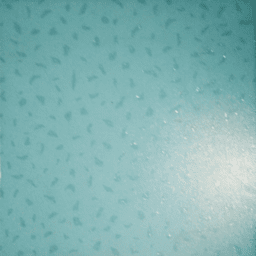} \vspace{1mm} \\
\end{tabular} & \hspace{-2.5mm} \includegraphics[align=c, width=0.135\linewidth]{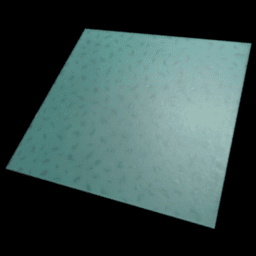} & \hspace{-3mm} \includegraphics[align=c, width=0.135\linewidth]{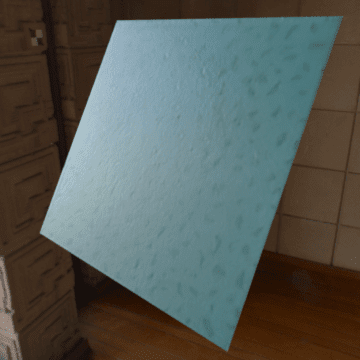} & \hspace{-3mm} \includegraphics[align=c, width=0.135\linewidth]{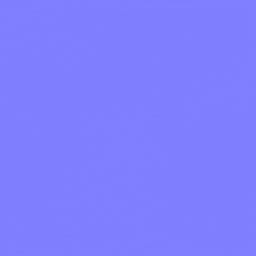} & \hspace{-3mm} \includegraphics[align=c, width=0.135\linewidth]{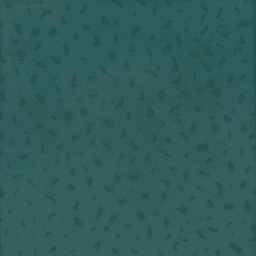} & \hspace{-3mm} \includegraphics[align=c, width=0.135\linewidth]{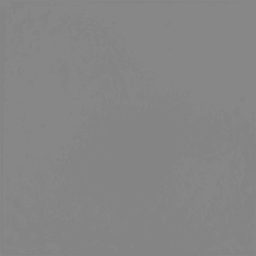} & \hspace{-3mm} \includegraphics[align=c, width=0.135\linewidth]{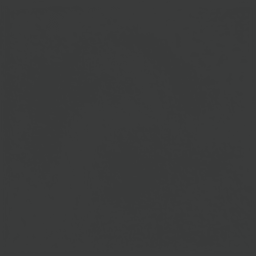} \vspace{3mm} \\
\begin{sideways} \hspace{-7mm} \tiny{Deschaintre et al. 2018} \end{sideways} & \hspace{-3mm} \includegraphics[align=c, width=0.135\linewidth]{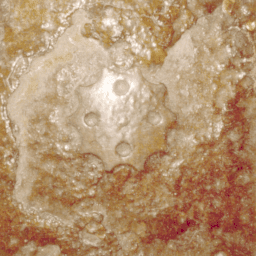} & \hspace{-2.5mm} \includegraphics[align=c, width=0.135\linewidth]{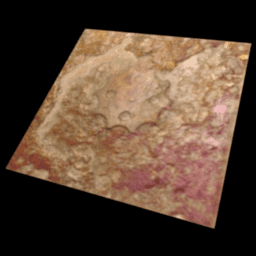} & \hspace{-3mm} \includegraphics[align=c, width=0.135\linewidth]{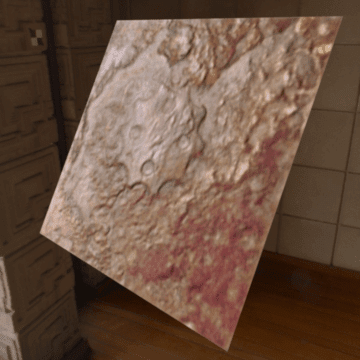} & \hspace{-3mm} \includegraphics[align=c, width=0.135\linewidth]{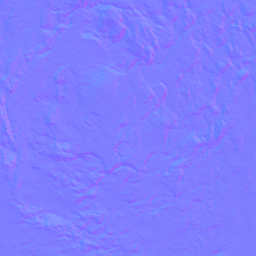} & \hspace{-3mm} \includegraphics[align=c, width=0.135\linewidth]{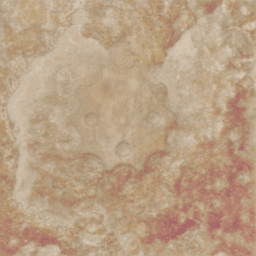} & \hspace{-3mm} \includegraphics[align=c, width=0.135\linewidth]{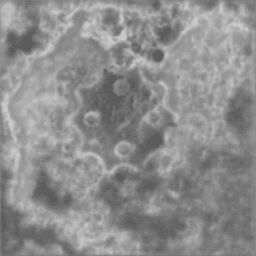} & \hspace{-3mm} \includegraphics[align=c, width=0.135\linewidth]{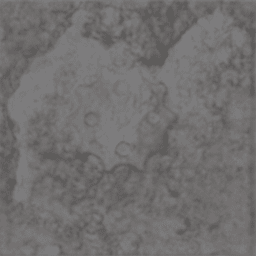} \vspace{1mm} \\
\begin{sideways} \hspace{-7mm} \tiny{Li et al. 2018} \end{sideways} & \hspace{-3mm} \includegraphics[align=c, width=0.135\linewidth]{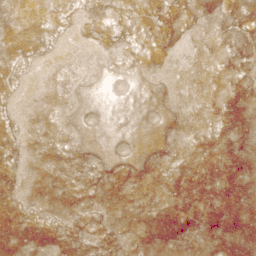} & \hspace{-2.5mm} \includegraphics[align=c, width=0.135\linewidth]{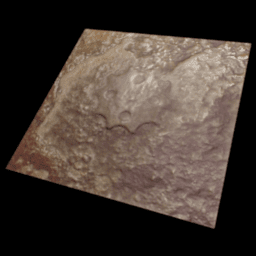} & \hspace{-3mm} \includegraphics[align=c, width=0.135\linewidth]{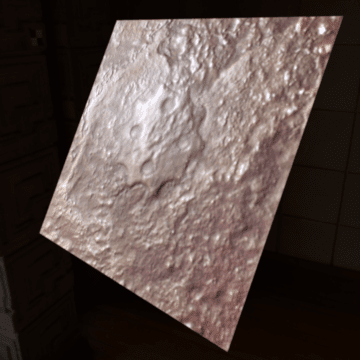} & \hspace{-3mm} \includegraphics[align=c, width=0.135\linewidth]{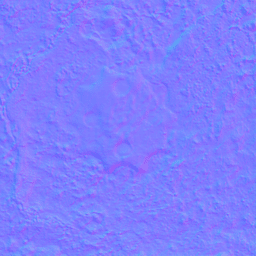} & \hspace{-3mm} \includegraphics[align=c, width=0.135\linewidth]{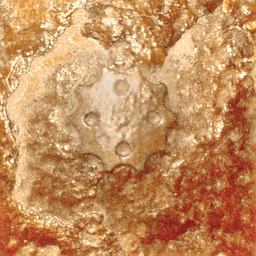} & \hspace{-3mm} \includegraphics[align=c, width=0.135\linewidth]{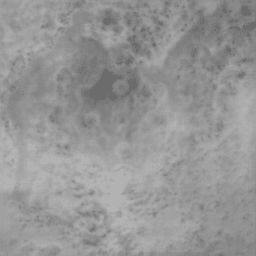} &   \vspace{1mm} \\
\begin{sideways} \hspace{-7mm} \small{Ours (4 inputs)} \end{sideways} & \begin{tabular} {cc}
\hspace{-2.5mm} \includegraphics[align=c, width=0.063\linewidth]{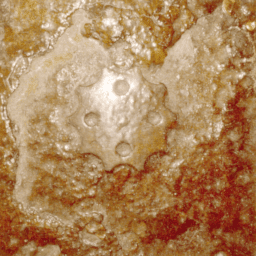} & \hspace{-3mm} \includegraphics[align=c, width=0.063\linewidth]{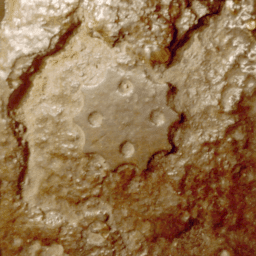} \vspace{1mm} \\
\hspace{-2.5mm} \includegraphics[align=c, width=0.063\linewidth]{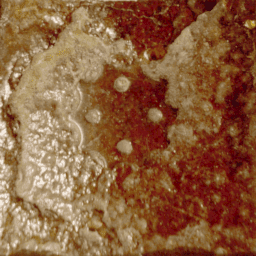} & \hspace{-3mm} \includegraphics[align=c, width=0.063\linewidth]{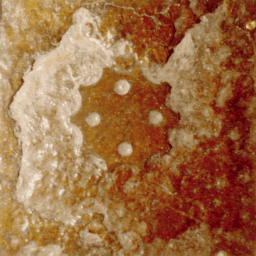} \vspace{1mm} \\
\end{tabular} & \hspace{-2.5mm} \includegraphics[align=c, width=0.135\linewidth]{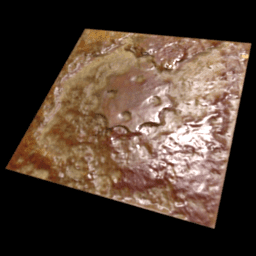} & \hspace{-3mm} \includegraphics[align=c, width=0.135\linewidth]{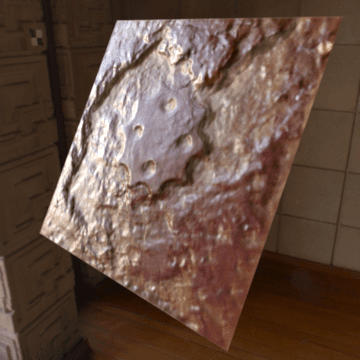} & \hspace{-3mm} \includegraphics[align=c, width=0.135\linewidth]{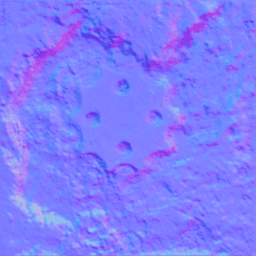} & \hspace{-3mm} \includegraphics[align=c, width=0.135\linewidth]{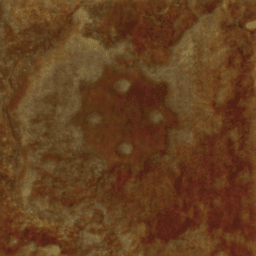} & \hspace{-3mm} \includegraphics[align=c, width=0.135\linewidth]{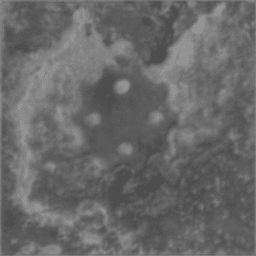} & \hspace{-3mm} \includegraphics[align=c, width=0.135\linewidth]{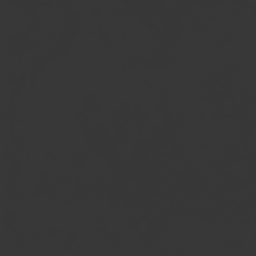} \vspace{1mm} \\
\end{tabular}

\caption{Comparison against single-image methods on real-world pictures. Our method recovers more normal details, and better removes highlight and shading residuals from the diffuse albedo. See supplemental materials for more comparisons and results.}
\label{fig:comparisonOneImageReal}
\end{figure*}

\begin{figure}[t]
\begin{tabular} {cccc}
\hspace{-5mm} \small{Deschaintre et al. 18} & 
\hspace{-4mm} \small{Li et al. 18} &
 \hspace{-3mm} \small{Ours (5 inputs)} & 
 \hspace{-2mm} \small{Ground truth} \\ 
 
 \vspace{1mm}
 
\hspace{-5mm} \includegraphics[align=c, width=0.25\linewidth]{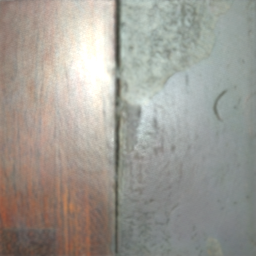} & 
\hspace{-4mm} \includegraphics[align=c, width=0.25\linewidth]{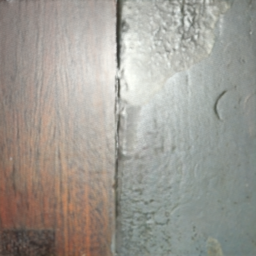} &
\hspace{-3mm}\includegraphics[align=c, width=0.25\linewidth]{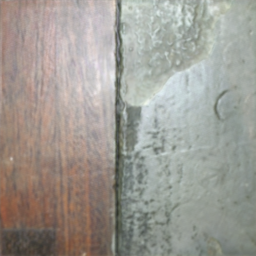} & 
\hspace{-2mm}\includegraphics[align=c, width=0.25\linewidth]{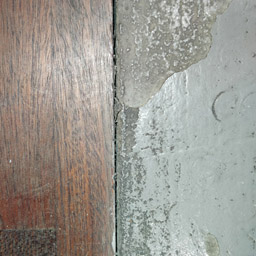}\\

\vspace{2mm}
 
\hspace{-4mm}\includegraphics[align=c, width=0.25\linewidth]{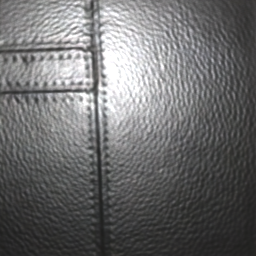} & 
\hspace{-3mm}\includegraphics[align=c, width=0.25\linewidth]{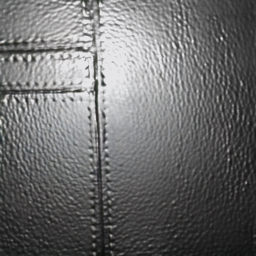} &
\hspace{-3mm}\includegraphics[align=c, width=0.25\linewidth]{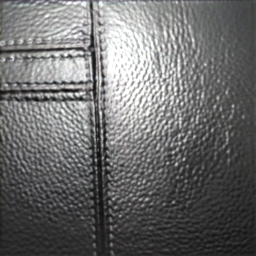} &
\hspace{-2mm}\includegraphics[align=c, width=0.25\linewidth]{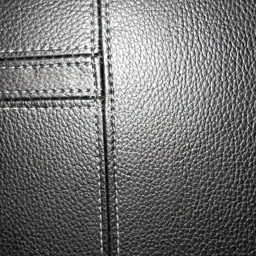}

\end{tabular}

\caption{ \NEW{Comparison to real-world relighting. Each column shows re-renderings of a captured material, except the last column which shows a picture of that material under a similar lighting condition (not used as input). We manually adjusted the position of the virtual light to best match the ground truth. Similarly, we adjusted the light power for each method separately since each has its own arbitrary scale factor.
Overall, our method better reproduces the normal and gloss variations of the materials. In particular, single-image methods tend to flatten the bumps of the leather and orient them towards the center of the picture, where the flash highlight appeared in the input. For individual result maps, see supplemental materials.
%Comparison of real materials acquired using different methods to a real picture. View and light conditions are manually replicated and light power is adapted for each method to provide the closest result to the picture.
}}
\label{fig:reRenderingsRealComparison}

\end{figure}

\begin{figure*}[!t]
\hspace{-7mm}\begin{tabular} {m{0.5em}cccccccc}
\rotatebox[origin=l]{90}{\tiny{Li et al.2018}} & \hspace{-3.5mm} \includegraphics[align=c, width=0.12\linewidth]{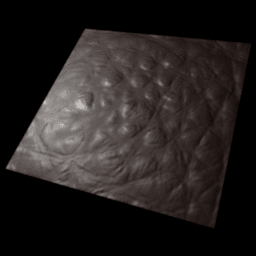} & \hspace{-3.5mm} \includegraphics[align=c, width=0.12\linewidth]{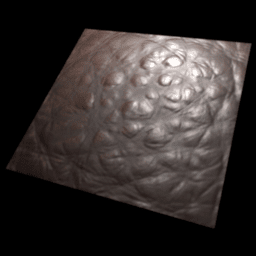} & \hspace{-3.5mm} \includegraphics[align=c, width=0.12\linewidth]{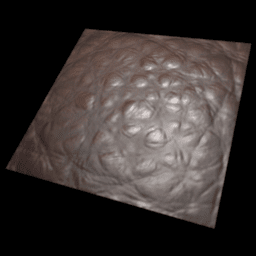} & \hspace{-3.5mm} \includegraphics[align=c, width=0.12\linewidth]{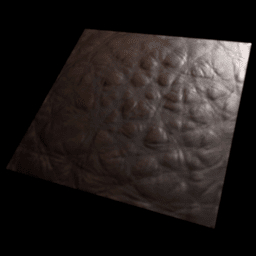} & \hspace{-1mm} \includegraphics[align=c, width=0.12\linewidth]{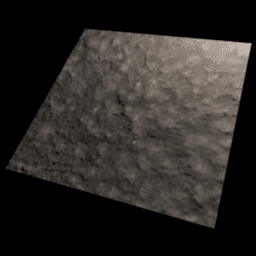} & \hspace{-3.5mm} \includegraphics[align=c, width=0.12\linewidth]{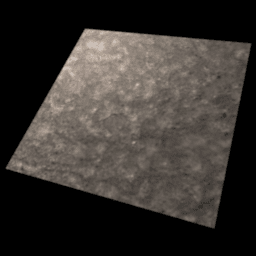} & \hspace{-3.5mm} \includegraphics[align=c, width=0.12\linewidth]{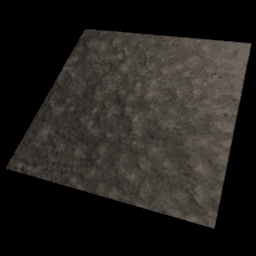} & \hspace{-3.5mm} \includegraphics[align=c, width=0.12\linewidth]{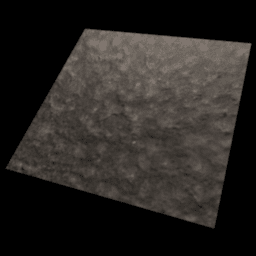} \vspace{1.5mm} \\
\rotatebox[origin=l]{90}{\tiny{Deschaintre et al.2018}}  & \hspace{-3.5mm} \includegraphics[align=c, width=0.12\linewidth]{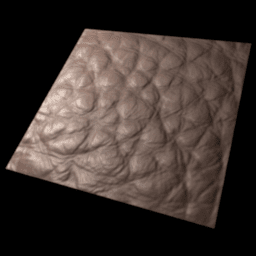} & \hspace{-3.5mm} \includegraphics[align=c, width=0.12\linewidth]{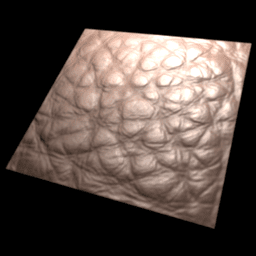} & \hspace{-3.5mm} \includegraphics[align=c, width=0.12\linewidth]{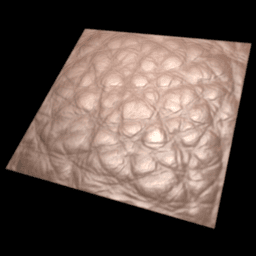} & \hspace{-3.5mm} \includegraphics[align=c, width=0.12\linewidth]{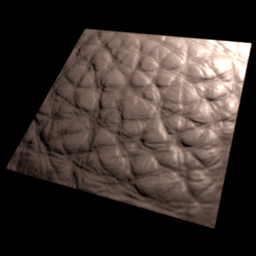} & \hspace{-1mm} \includegraphics[align=c, width=0.12\linewidth]{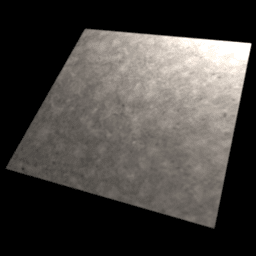} & \hspace{-3.5mm} \includegraphics[align=c, width=0.12\linewidth]{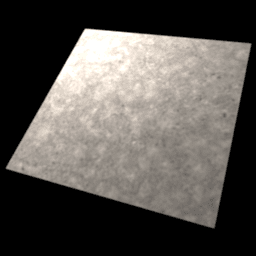} & \hspace{-3.5mm} \includegraphics[align=c, width=0.12\linewidth]{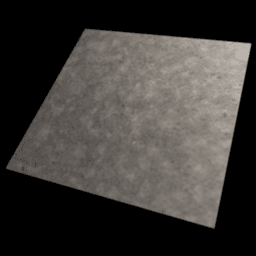} & \hspace{-3.5mm} \includegraphics[align=c, width=0.12\linewidth]{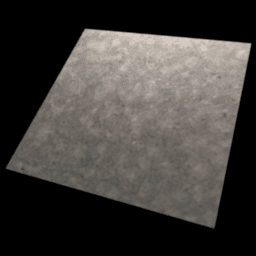} \vspace{1.5mm} \\
\rotatebox[origin=l]{90}{\small{BTF}}  & \hspace{-3.5mm} \includegraphics[align=c, width=0.12\linewidth]{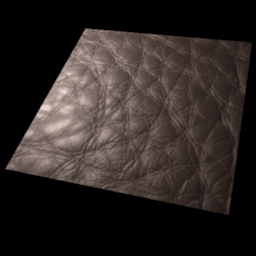} & \hspace{-3.5mm} \includegraphics[align=c, width=0.12\linewidth]{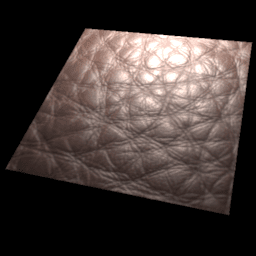} & \hspace{-3.5mm} \includegraphics[align=c, width=0.12\linewidth]{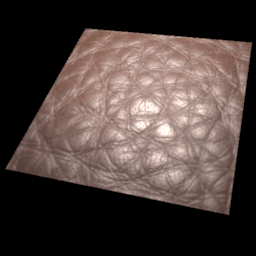} & \hspace{-3.5mm} \includegraphics[align=c, width=0.12\linewidth]{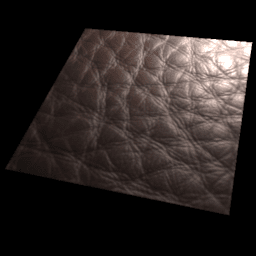} & \hspace{-1mm} \includegraphics[align=c, width=0.12\linewidth]{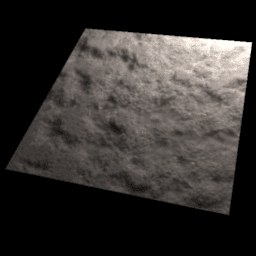} & \hspace{-3.5mm} \includegraphics[align=c, width=0.12\linewidth]{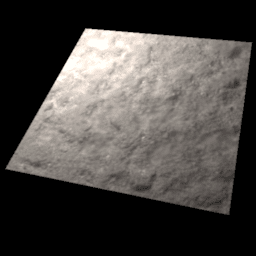} & \hspace{-3.5mm} \includegraphics[align=c, width=0.12\linewidth]{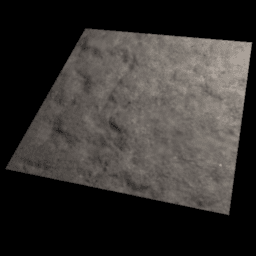} & \hspace{-3.5mm} \includegraphics[align=c, width=0.12\linewidth]{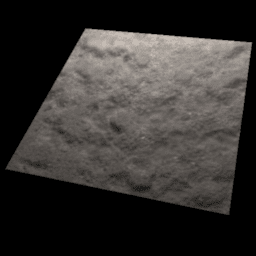} \vspace{1.5mm} \\
\rotatebox[origin=l]{90}{\small{Ours (10 inputs)}}  & \hspace{-3.5mm} \includegraphics[align=c, width=0.12\linewidth]{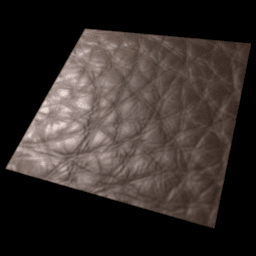} & \hspace{-3.5mm} \includegraphics[align=c, width=0.12\linewidth]{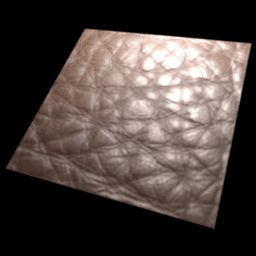} & \hspace{-3.5mm} \includegraphics[align=c, width=0.12\linewidth]{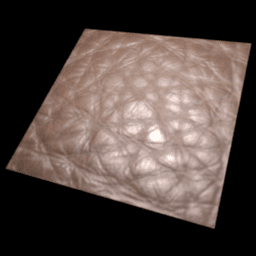} & \hspace{-3.5mm} \includegraphics[align=c, width=0.12\linewidth]{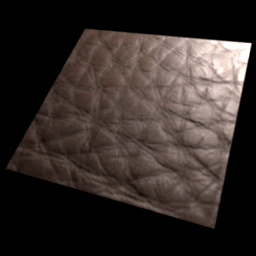} & \hspace{-1mm} \includegraphics[align=c, width=0.12\linewidth]{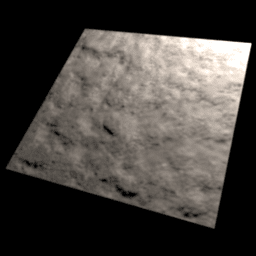} & \hspace{-3.5mm} \includegraphics[align=c, width=0.12\linewidth]{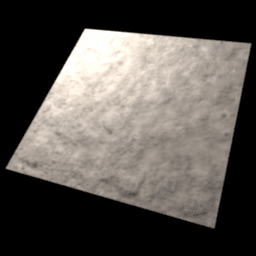} & \hspace{-3.5mm} \includegraphics[align=c, width=0.12\linewidth]{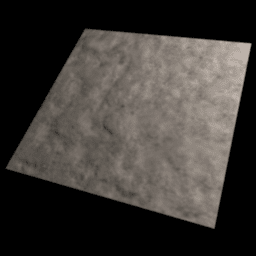} & \hspace{-3.5mm} \includegraphics[align=c, width=0.12\linewidth]{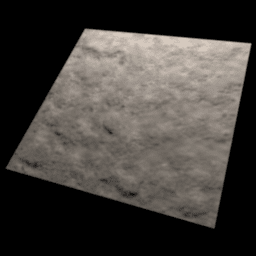} \vspace{1.5mm} \\
\end{tabular}
\caption{Comparison against single-image methods on a measured BTF with ground truth re-renderings. Our method globally captures the material features better.}
\label{fig:bonnComparisonRenderings}
\end{figure*}

We next compare to the recent single-image methods of Deschaintre et al.~\shortcite{Deschaintre18} and Li et al.~\shortcite{Li18}, which both take as input a fronto-parallel flash photo. Figure~\ref{fig:comparisonOneImageSynth} provides a visual comparison on synthetic SVBRDFs with ground truth maps, 
Figure~\ref{fig:bonnComparisonRenderings} provides a similar comparison on BTFs measured from 81x81 pictures, which allow ground-truth re-renderings, and Figure~\ref{fig:comparisonOneImageReal} \NEW{ and} \ref{fig:reRenderingsRealComparison} provide a comparison on real pictures. While developed concurrently, both single-image approaches suffer from the same limitations. The co-located lighting tends to produce low-contrast shading, reducing the cues available for the network to fully retrieve normals. Adding side-lit pictures of the material helps our approach retrieve these missing details. The fronto-parallel flash also often produces a saturated highlight in the middle of the image, which both single-image methods struggle to in-paint convincingly in the different maps. While the strength of the highlight could be reduced by careful tuning of exposure, saturated pixels are difficult to avoid in real-world capture. In contrast, our method benefits from additional pictures to recover information about those pixels.

%We provide several comparisons with two previous single-photo methods that use a fronto-parallel flash photo, namely Deschaintre et al.~\shortcite{Deschaintre18} and Li et al.~\shortcite{Li18}. Using a fronto-parallel flash-picture setup causes the shadows to be squashed, reducing the cues available for the network to fully retrieve normals. Adding side-lit pictures of the material provides more information about the shadowing effect created by strong normals. \TODO{normalSquash figure ?}
%, as shown in fig.\ref{fig:normalSquash}. 

%We first show synthetic materials in Fig.~\ref{fig:fig:comparisonOneImageSynth}, where our method is clearly closer to ground truth for the consituent maps. The additional information provided by multiple lighting conditions allows the network do disambiguate the intricate details of each map.

%Similarly, comparisons on real data Fig.~\ref{fig:comparisonOneImageReal} qualitatively show that our method improves the estimation of the SVBRDF. This becomes evident in the re-renderings, where the appearance of the captured material is better preserved. Again, the specular ``splotch'' is typically removed, and more accurate normals are captured, while a more plausible specularity/roughness is often retrieved. 

Another limitation of these two single-image methods is that the flash highlight cannot cover all parts of the material sample. This lack of information can cause erroneous estimations, especially when the sample is composed of multiple materials with different shininess. 
%will often show on one material of the whole SVBRDF. This will allow the method to retrieve plausible parameters for this material, but will lack information for any other material in the picture. 
Providing more pictures gives a chance to our method to observe highlights over all parts of the sample, as is the case in Figure~\ref{fig:multipleMaterials}, where the difference in roughness in the upper right only becomes apparent with the 4th input.

\begin{figure*}[!t]
\hspace{-8mm} \begin{tabular} {cccccccc}
  & \small{Inputs} & \multicolumn{2}{c}{\small{Renderings}} & \small{Normal} & \small{Diffuse} & \small{Roughness} & \small{Specular}
 \vspace{1mm} \\
\begin{sideways} \hspace{-5mm} \small{1 input} \end{sideways} & \hspace{-3mm} \includegraphics[align=c, width=0.14\linewidth]{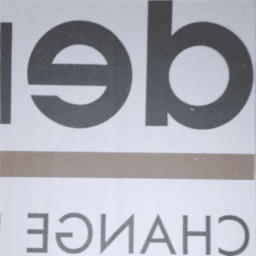} & \hspace{-1mm} \includegraphics[align=c, width=0.14\linewidth]{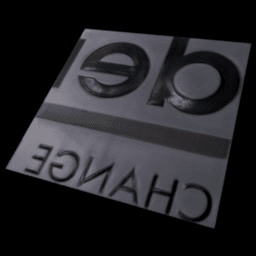} & \hspace{-3mm} \includegraphics[align=c, width=0.14\linewidth]{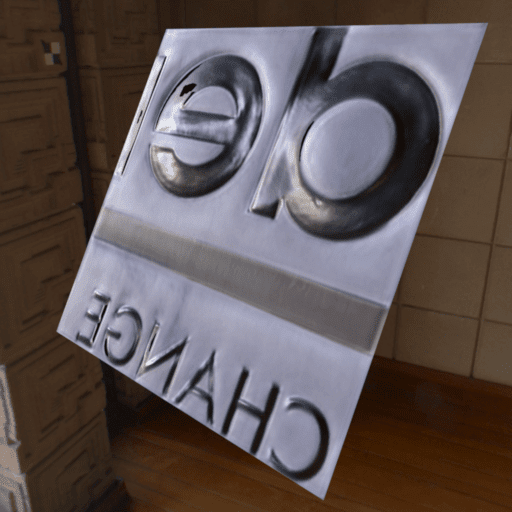} & \hspace{-3mm} \includegraphics[align=c, width=0.14\linewidth]{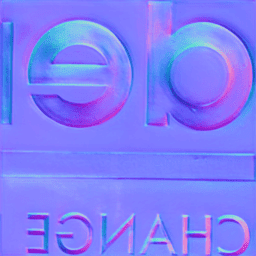} & \hspace{-3mm} \includegraphics[align=c, width=0.14\linewidth]{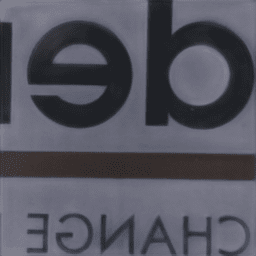} & \hspace{-3mm} \includegraphics[align=c, width=0.14\linewidth]{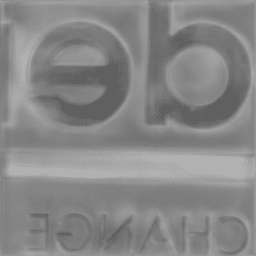} & \hspace{-3mm} \includegraphics[align=c, width=0.14\linewidth]{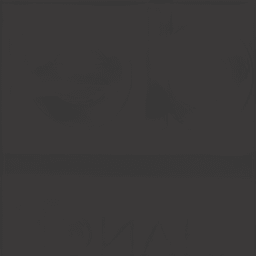} \vspace{1mm} \\
\begin{sideways} \hspace{-5mm} \small{2 inputs} \end{sideways} & \hspace{-3mm} \includegraphics[align=c, width=0.14\linewidth]{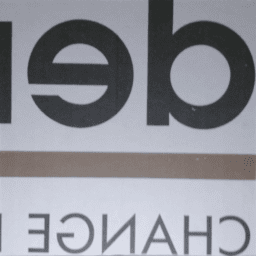} & \hspace{-1mm} \includegraphics[align=c, width=0.14\linewidth]{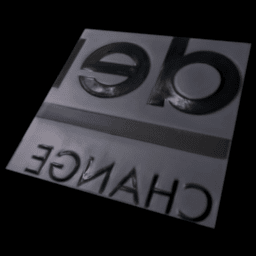} & \hspace{-3mm} \includegraphics[align=c, width=0.14\linewidth]{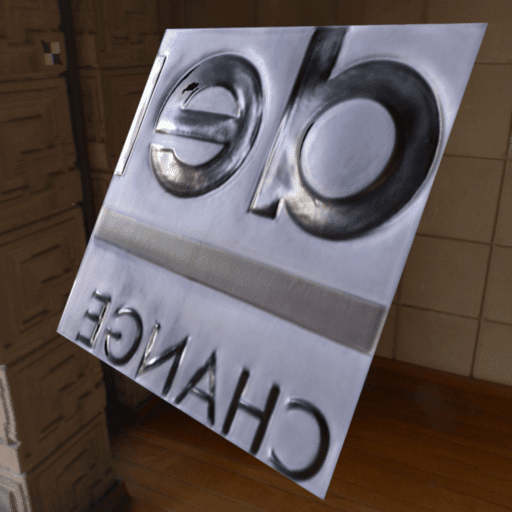} & \hspace{-3mm} \includegraphics[align=c, width=0.14\linewidth]{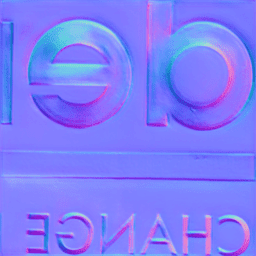} & \hspace{-3mm} \includegraphics[align=c, width=0.14\linewidth]{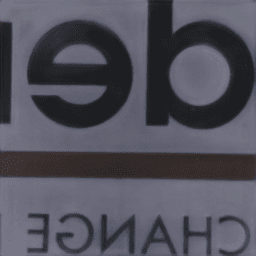} & \hspace{-3mm} \includegraphics[align=c, width=0.14\linewidth]{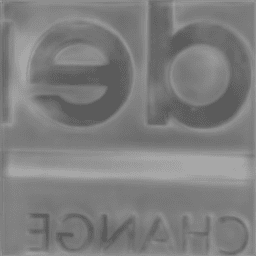} & \hspace{-3mm} \includegraphics[align=c, width=0.14\linewidth]{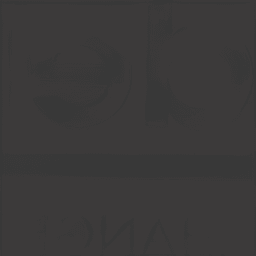} \vspace{1mm} \\
\begin{sideways} \hspace{-5mm} \small{4 inputs} \end{sideways} & \hspace{-3mm} \includegraphics[align=c, width=0.14\linewidth]{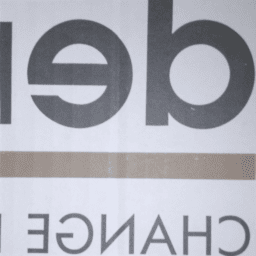} & \hspace{-1mm} \includegraphics[align=c, width=0.14\linewidth]{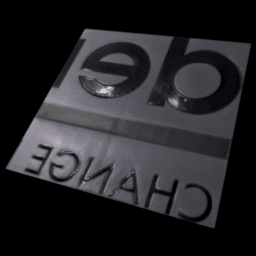} & \hspace{-3mm} \includegraphics[align=c, width=0.14\linewidth]{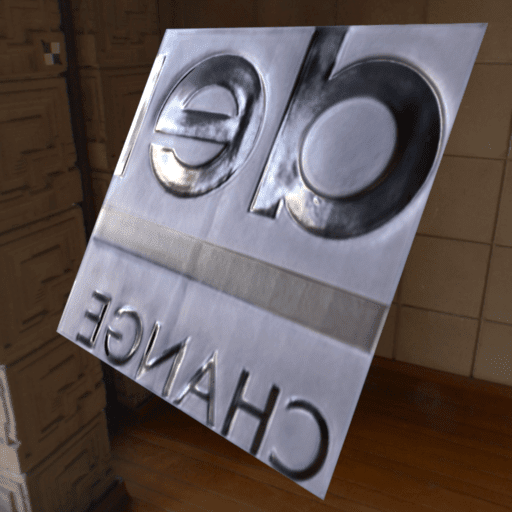} & \hspace{-3mm} \includegraphics[align=c, width=0.14\linewidth]{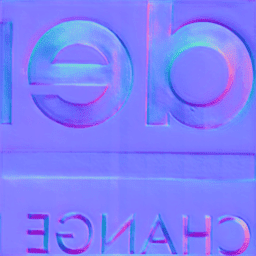} & \hspace{-3mm} \includegraphics[align=c, width=0.14\linewidth]{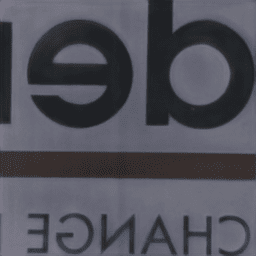} & \hspace{-3mm} \includegraphics[align=c, width=0.14\linewidth]{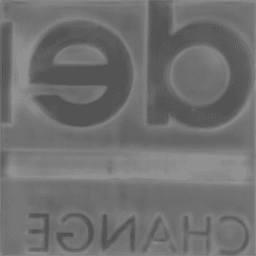} & \hspace{-3mm} \includegraphics[align=c, width=0.14\linewidth]{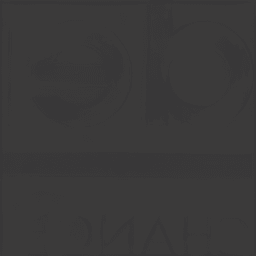} \vspace{3mm} \\
\begin{sideways} \hspace{-5mm} \small{1 input} \end{sideways} & \hspace{-3mm} \includegraphics[align=c, width=0.14\linewidth]{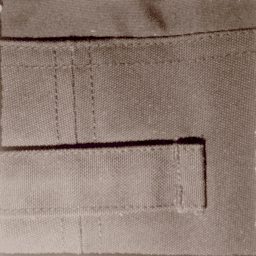} & \hspace{-1mm} \includegraphics[align=c, width=0.14\linewidth]{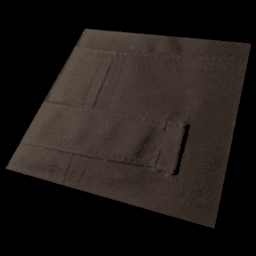} & \hspace{-3mm} \includegraphics[align=c, width=0.14\linewidth]{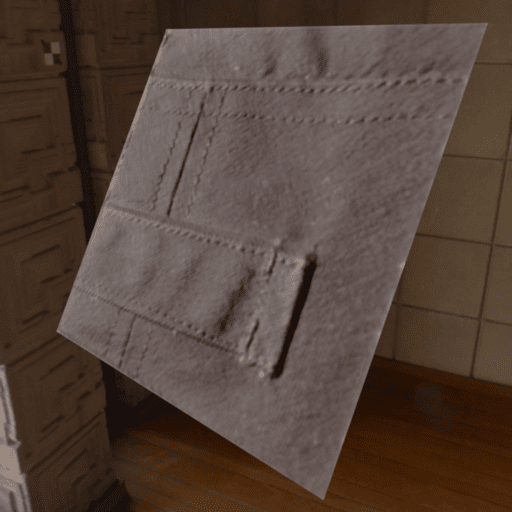} & \hspace{-3mm} \includegraphics[align=c, width=0.14\linewidth]{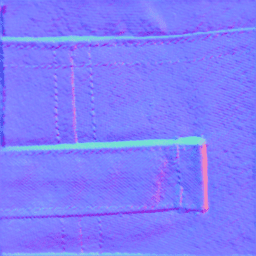} & \hspace{-3mm} \includegraphics[align=c, width=0.14\linewidth]{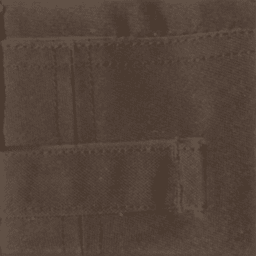} & \hspace{-3mm} \includegraphics[align=c, width=0.14\linewidth]{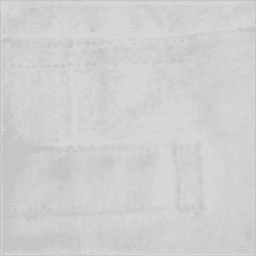} & \hspace{-3mm} \includegraphics[align=c, width=0.14\linewidth]{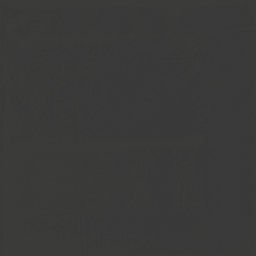} \vspace{1mm} \\
\begin{sideways} \hspace{-5mm} \small{2 inputs} \end{sideways} & \hspace{-3mm} \includegraphics[align=c, width=0.14\linewidth]{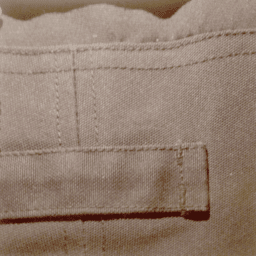} & \hspace{-1mm} \includegraphics[align=c, width=0.14\linewidth]{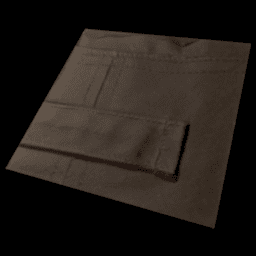} & \hspace{-3mm} \includegraphics[align=c, width=0.14\linewidth]{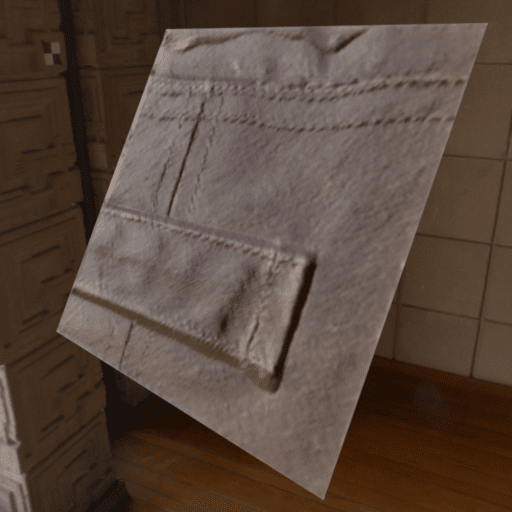} & \hspace{-3mm} \includegraphics[align=c, width=0.14\linewidth]{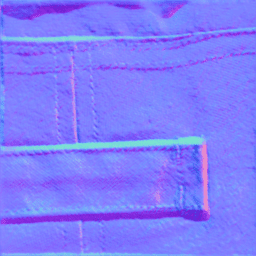} & \hspace{-3mm} \includegraphics[align=c, width=0.14\linewidth]{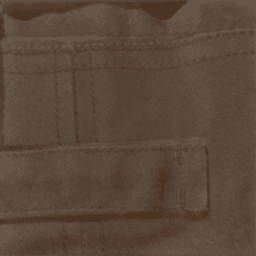} & \hspace{-3mm} \includegraphics[align=c, width=0.14\linewidth]{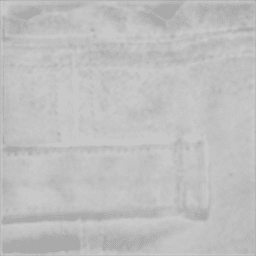} & \hspace{-3mm} \includegraphics[align=c, width=0.14\linewidth]{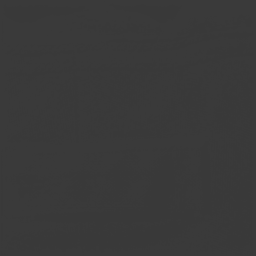} \vspace{1mm} \\
\begin{sideways} \hspace{-5mm} \small{4 inputs} \end{sideways} & \hspace{-3mm} \includegraphics[align=c, width=0.14\linewidth]{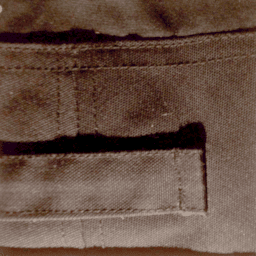} & \hspace{-1mm} \includegraphics[align=c, width=0.14\linewidth]{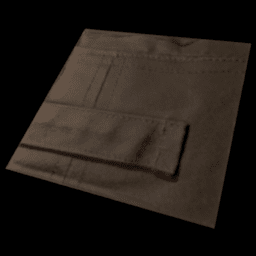} & \hspace{-3mm} \includegraphics[align=c, width=0.14\linewidth]{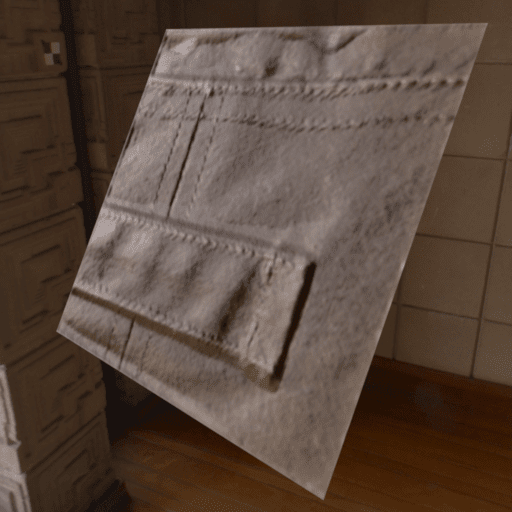} & \hspace{-3mm} \includegraphics[align=c, width=0.14\linewidth]{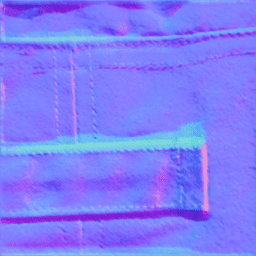} & \hspace{-3mm} \includegraphics[align=c, width=0.14\linewidth]{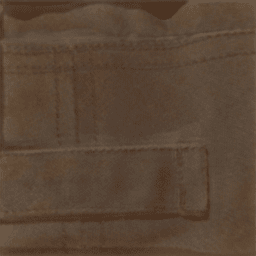} & \hspace{-3mm} \includegraphics[align=c, width=0.14\linewidth]{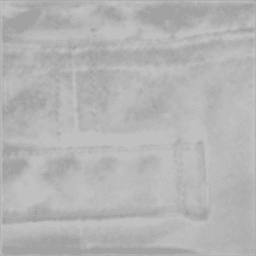} & \hspace{-3mm} \includegraphics[align=c, width=0.14\linewidth]{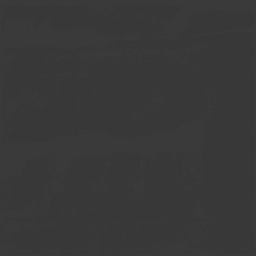} \vspace{1mm} \\
\end{tabular}

\caption{Limitations. We inherits some of the limitations of the method by Deschaintre et al. \protect{\shortcite{Deschaintre18}}, such as the tendency to produce correlated maps and to interpret dark pixels as shiny (top). Our SVBRDF representation, training data and loss do not model cast shadows. As a result, shadows in the input pollute some of the maps (bottom).}
\label{fig:limitations}
\end{figure*}

\subsection{Limitations}
Since our method builds on the single-image network of Deschaintre et al.~\shortcite{Deschaintre18}, it inherits some of its limitations. First, the method is limited to materials that can be well represented by an isotropic Cook-Torrance BRDF. We also observe that the method tends to produce correlated maps and interpret dark materials as shiny, as shown in Figure~\ref{fig:limitations}(top) where despite several pictures, albedo variations of the cardboard get interpreted as normal variations, and the black letters get assigned a low roughness. This behavior reflects the content of our training data, since most artist-designed SVBRDFs have correlated maps.
%Finally, the method sometimes has difficulty recovering the roughness of very smooth materials, as shown with the Scotch tape in Figure~\ref{fig:limitation}. This may be due to the relatively limited impact that concentrated specular highlights have on the rendering loss.

Since we rectify the multi-view inputs with a simple homography, we do not correct for parallax effects produced by surfaces with high relief. This approximation may yield misalignment in the input images, which in turn reduces the sharpness of the predicted maps. In addition, our SVBRDF representation, training data, and rendering loss do not model cast shadows. While shadows are mostly absent in pictures taken with a co-located flash, they can appear when using a handheld flash and remain visible in some of our results, as shown in Figure~\ref{fig:limitations} (bottom).

%Finally, we observed that the method tends to slightly over-estimate the roughness of very smooth materials, and overall has difficulty improving roughness estimation given additional inputs (Figure~\ref{fig:globalSSIM}). As a result, while our method provides a very good SVBRDF estimation from a few pictures, it only improves marginally after 10 inputs and does not reach yet the quality obtained by a classical optimization when dozens of images are available (Figure~\ref{fig:classicalOptim}). This staturation may be due to the limited impact that concentrated specular highlights have on the rendering loss. An interesting direction for future work would be to implement differentiable rendering of more complex light sources, such as area lights \cite{Heitz2016} and environment maps. This would not only increase the flexibility of the method in terms of capture, it might also improve its performance since extended light sources are known to better convey gloss than point lights \cite{Fleming03}.

\section{Conclusion}
With the advance of deep learning, the holy grail of single-image SVBRDF capture recently became a reality. Yet, despite impressive results, single-image methods offer little margin to users to correct for erroneous predictions. We address this fundamental limitation with a deep network architecture that accepts a variable number of input images, allowing users to capture as many images as needed to exhibit all the visual effects they want to capture of a material. Our method bridges the gap between single-image and many-image methods, allowing faithful material capture with a handful of images captured from \REM{uncontrolled,} uncalibrated light-view directions.

\section*{Acknowledgments}
% Who do we want to thank ?
We thank Yulia Gryaditskaya, Simon Rodriguez and Stavros Diolatzis for their support during the deadline as well as Anthony Jouanin and Vincent Hourdin for regular feedback. We also thank Zhengqin Li and Kalyan Sunkavalli for their help with evaluation.
This work was partially funded by an ANRT (http://www.anrt.asso.fr/en) CIFRE scholarship between Inria and Optis, the ERC Advanced Grant FUNGRAPH (No. 788065, http://fungraph.inria.fr), and by software and hardware donations from Adobe and Nvidia.

\bibliographystyle{eg-alpha-doi} 
\bibliography{bibtex/bibliography} 
\end{document}